\documentclass[sigconf,nonacm]{acmart}

\usepackage{multirow}
\usepackage{caption}
\usepackage{listings}
\usepackage{xcolor}

\lstdefinelanguage{none}{
  keywords={},
  keywordstyle=\color{black},
  sensitive=false,
  comment=[l]{//},
  morecomment=[l]{//}
}

\lstdefinestyle{promptblock}{
  language=none,
  basicstyle=\ttfamily\scriptsize,
  breaklines=true,
  frame=single,
  backgroundcolor=\color{gray!5},
  captionpos=b
}

\AtBeginDocument{%
  }

\renewcommand\footnotetextcopyrightpermission[1]{}
\date{Preprint. Under review.}

\begin{document}

%%
%% The ''title'' command has an optional parameter,
%% allowing the author to define a ''short title'' to be used in page headers.
\newcommand{\todo}[1]{\textcolor{blue}{#1}}
\newcommand{\old}[1]{\textcolor{purple}{#1}}
\newcommand{\systemname}{Imago Obscura}

\title[\systemname: An Image Privacy AI Co-pilot]{\systemname: An Image Privacy AI Co-pilot to Enable Identification and Mitigation of Risks}

\author{Kyzyl Monteiro}
\authornotemark[1]
\email{kyzyl@cmu.edu}
\orcid{0000-0002-2723-9500}
\affiliation{%
  \institution{Carnegie Mellon University}
  \city{Pittsburgh}
  \state{PA}
  \country{USA}
}

\author{Yuchen Wu}
\authornote{Both authors contributed equally to this work.}
\email{wuyuchen21@mails.tsinghua.edu.cn}
\orcid{0009-0008-7045-4477}
\affiliation{%
  \institution{Tsinghua University}
  \city{Beijing}
  \country{China}
}
\author{Sauvik Das}
\email{sauvik@cmu.edu}
\orcid{0000-0002-9073-8054}
\affiliation{%
  \institution{Carnegie Mellon University}
  \city{Pittsburgh}
  \state{PA}
  \country{USA}
}

\renewcommand{\shortauthors}{Monteiro et al.}

\begin{abstract}
Users often struggle to navigate the privacy / publicity boundary in sharing images online: they may lack awareness of image privacy risks and/or the ability to apply effective mitigation strategies. To address this challenge, we introduce and evaluate \systemname, an AI-powered, image-editing copilot that enables users to identify and mitigate privacy risks with images they intend to share.
Driven by design requirements from a formative user study with 7 image-editing experts, \systemname{} enables users to articulate their image-sharing intent and privacy concerns.
The system uses these inputs to surface contextually pertinent privacy risks, and then recommends and facilitates application of a suite of obfuscation techniques found to be effective in prior literature --- e.g., inpainting, blurring, and generative content replacement. We evaluated \systemname{} with 15 end-users in a lab study and found that it greatly improved users' awareness of image privacy risks and their ability to address those risks, allowing them to make more informed sharing decisions.
 
\end{abstract}




\begin{teaserfigure}
  \includegraphics[width=\textwidth]{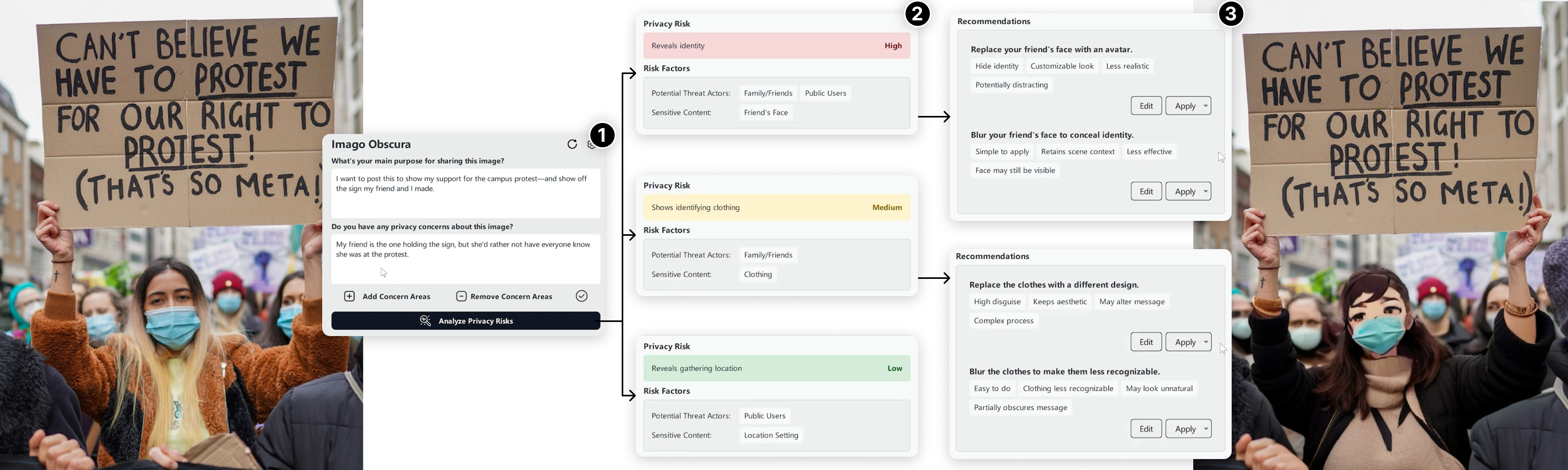}
  \vspace{-17pt}
  \caption{\systemname: A privacy-focused image AI-copilot that enables users to: 1) articulate their image sharing intent and privacy concerns; 2) become aware of multiple contextually pertinent image privacy risks; and 3) apply recommended obfuscation techniques for the risks they choose to address, enabling informed decision-making about image sharing.}
  \Description{A sequence of interface panels showing how a protest photo is analyzed and obfuscated by Imago Obscura. The user enters sharing intent and privacy concerns; the system identifies face, clothing, and location risks; recommends obfuscation strategies; and applies edits, replacing a face with a generative avatar and clothing with a different set of clothing}
  \label{fig:teaser}
\end{teaserfigure}

\maketitle

\begin{figure*}
  \includegraphics[width=\textwidth]{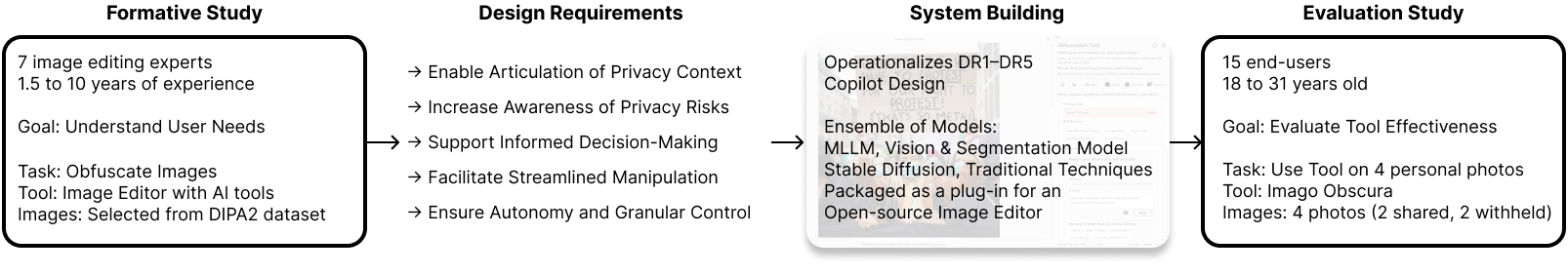}
  \vspace{-25pt}
  \caption{Overview of our methodology. We conducted a formative study to derive design requirements, built a tool based on those requirements, and evaluated it with end-users using their personal photos.}
  \Description{The figure outlines the study methodology: it begins with a formative study where image editing experts obfuscated images to surface user needs. From this, five design requirements were derived, including supporting informed decision-making and enabling granular control. These informed the development of a system combining multiple AI models into an open-source image editor. The final stage was an evaluation study where end-users used the tool on personal photos.}
  \label{fig:method}
\end{figure*}
  \vspace{-25pt}
\section{Introduction}
One of the greatest usable privacy challenges of the modern social internet is helping users navigate what Palen and Dourish, in 2003, identified as the `privacy/publicity' boundary: i.e., people's desire to share personal information with others without exposing themselves to undue risks \cite{palen2003unpacking}.
This problem manifests acutely in the context of image sharing --- people collectively share 14 billion images daily \cite{broz2024how}, for reasons ranging from sharing and documenting moments in their personal lives to collaborating and communicating in their professional lives \cite{castro2015predicting, van2005uses}. But sharing images also comes with risks: many personal images can reveal a wide range of potentially sensitive information: e.g., who one knows, where one goes, and what one likes to do
\cite{ahern2007over, fire2014online, patsakis2015privacy}.
These risks are not abstract: prior work has shown that violations of image privacy and security can lead to a spectrum of harms, including embarrassment, job loss, identity theft, stalking, and harassment \cite{rashidi2018you, wang2011regretted, mantouvalou2019lost, vishwamitra2021towards}.

Despite these risks, a large body of prior art suggests that many users have trouble understanding and mitigating the privacy risks associated with sharing personal images online \cite{henne2013awareness, kumar2013security, nyoni2018privacy, wang2011regretted, li2022obfuscation, liu2022privacy}.
Some users are \textit{\textbf{unaware}} of the risks they expose themselves to inadvertently while sharing the image \cite{henne2013awareness, kumar2013security, nyoni2018privacy, wang2011regretted}; others may find they have \textbf{\textit{limited ability}} to mitigate these risks because existing approaches to obfuscate or edit-out sensitive information from images require significant technical expertise \cite{li2022obfuscation, liu2022privacy}.
As a result, today, users generally employ crude and restrictive strategies when navigating the privacy/publicity boundary for sharing images online --- i.e., they ignore privacy risks altogether, self-censor themselves, or utilize insufficient and error-prone audience selection controls to try and limit who can see their images \cite{sleeper2013post, fogues2017exploring, li2019hideme, zerr2012picalert}.

How can we make it easier for users to effectively identify and mitigate privacy risks in images they want to share online?
To answer that question, we introduce \systemname{}---an image privacy copilot that leverages generative AI technologies to help end-users identify image privacy risks, mitigate those risks, and make more informed decisions about what images to share online. We employed a three-phased, human-centered design process to design and evaluate \systemname{}.

First, we conducted \textbf{a formative study} with seven image-editing experts to derive a deeper understanding of user requirements for an image privacy protection tool. From this study, we distilled five design requirements to prioritize when developing \systemname{}. For example, we learned that while there is a general need to raise users' awareness of image privacy risks, it is imperative to do this in a manner that is customized to users' \textit{lived concerns and contexts of use}. We also learned that there is a need to \textit{facilitate application of a broad range of image obfuscation techniques}, so that even users without sophisticated image editing skills can make an informed decision about how to obfuscate their image in a manner that addresses their concerns but does not significantly impede their sharing intent.

Second, we \textbf{designed and developed \systemname{}} based on the design requirements we derived from our formative study. \systemname{} enables users to directly articulate their privacy concerns and sharing intent through natural language, and uses this information to identify pertinent risks in users' images. It then recommends appropriate obfuscation techniques while informing users of the implications of those techniques, and automatically applies obfuscation strategies users choose to implement. Our system comprises an ensemble of AI models: a vision model that identifies and annotates objects in users' images \cite{xiao2023florence2advancingunifiedrepresentation}, a multimodal large language model \cite{openai2023gpt} that identifies pertinent risks, a segmentation model \cite{kirillov2023segment} and an image generation model \cite{karras2020training} that automatically and precisely applies obfuscation techniques. This ensemble is integrated into an open-source image editing tool \cite{krita}. Moreover, to ensure that \systemname's outputs are accurate and robust, we employ a theory-grounded prompting approach that scaffolds the AI output based on known understandings of image privacy risks \cite{ahern2007over, fire2014online, patsakis2015privacy} and image obfuscation techniques \cite{li2017effectiveness, hasan2018viewer, xu2024examining, khamis2022deepfakes} from the usable privacy literature.

Finally, we \textbf{evaluated \systemname{}} through a lab user study with 15 participants. In short, we found that \systemname{} adhered well to the core design requirements we identified in our formative study: participants found that the tool helped them mitigate the privacy concerns they really cared about, surfaced relevant risks that would have otherwise gone unnoticed, and aided them in making an informed decision about which risks to accept and which to mitigate based on their sharing intent. We also found some opportunities for improvement. For example, there is a need for guardrails to prevent malicious use --- as by lowering the barrier to applying obfuscation techniques to their images, \systemname{} also facilitates the creation of inauthentic or misleading images.
On balance, however, we found that \systemname{} helps users
make more informed decisions balancing their desires for privacy and publicity when sharing images online. 

In summary, this paper contributes:
\begin{enumerate}
    \item Five core design requirements for an image privacy copilot to help users make informed decisions about how to balance privacy concerns with sharing intent when obfuscating images they hope to share online.
    \item The design and implementation of a system, \textit{\systemname}, which demonstrates the following novel approaches in a Human-AI teaming effort for image privacy protection: 
    \begin{enumerate}
        \item Enabling users to express privacy concerns in natural language and receive tailored image privacy support. 
        \item An ensemble of models, grounded in prior literature, that automatically identifies both user-specified and overlooked image privacy risks. 
        \item Informed recommendations and one-click application of risk mitigation strategies through semantic image obfuscation techniques. 
    \end{enumerate}
    \item Insights from a user study which demonstrates participants appreciated a) the ability to articulate concerns as it helped them address risks they deemed most important; b) the comprehensive presentation of risks and recommended obfuscations, which supported informed decision-making; and c) the ability to apply recommended image obfuscations easily with precise control and high agency.
\end{enumerate}

\section{Threat Model}
Our adversary seeks to learn or infer sensitive personal information about the subject/owner of a photo, based on the content in the photo, that is peripheral to the sharing intent. To that end, we focus on \textit{observable privacy} and \textit{inferential privacy} \cite{liu2022privacy}, covering both visible and inferable content-based risks. We do not consider \textit{contextual privacy} \cite{liu2022privacy}, i.e. threats from captions or metadata and exclude threats from automated adversaries without an analyst-in-the-loop. Consequently, we do not implement defenses against algorithmic adversaries (e.g., adversarial perturbations \cite{shan2020fawkes,shan2023glaze,shan2023prompt}). Use case scenarios aligned with this model are provided in Appendix~\ref{sec:usecase}.

\section{Related Work}

\subsection{Identifying and Taxonomizing Privacy-Sensitive Content}

\paragraph{Studies and Taxonomies on Sensitive Content} A large body of work has focused on studying, identifying, and categorizing sensitive content in images. Through interview-based studies, researchers have highlighted various sensitive content elements such as faces, objects, backgrounds, and phone screens \cite{ahern2007over, hoyle2015sensitive, aura2006scanning}. Researchers have also identified, categorized, and studied sensitive content \cite{li2018human}, culminating in a taxonomy of 28 categories of sensitive content that was derived from prior literature and an analysis of photos collected from participants \cite{li2020towards}.

\paragraph{Datasets on Sensitive Content}
The computer vision and machine learning communities have also shown significant interest in this domain, resulting in the creation of various datasets. Some studies classify images as private or public \cite{zerr2012privacy, spyromitros2016personalized}, while others, more aligned with our holistic approach, categorized more granular categories of potentially private content like nudity, violence, and drinking \cite{zhao2022privacyalert, yu2016iprivacy}. Prior work in this domain include Yu et al.'s 268 privacy-sensitive object classes \cite{yu2017privacy, yu2018leveraging} and datasets like VISPR and PrivacyAlert \cite{orekondy2017towards, zhao2022privacyalert,orekondy2017towards}. Recent work by Xu et al. adds detailed reasoning on privacy perceptions in addition to object-level annotations \cite{xu2023dipa, xu2024dipa2}, which is important for content-level image privacy protection.

While the research community has developed a comprehensive understanding of image privacy, this awareness of privacy risks appears to be largely confined to academia, with a notable lack of awareness among end-users. Consequently, our work aims to distill the rich literature on usable image privacy into a practical and user-friendly tool that enhances end-users' awareness and ability to identify image privacy risks. We propose a user-facing pipeline that identifies sensitive objects and raises awareness of privacy risks.

\subsection{Image Obfuscation Techniques and Their Effectiveness}
Image obfuscation techniques have evolved from traditional methods like blurring and pixelation \cite{demanet2007pixelating, korshunov2013framework, lander2001evaluating} to advanced approaches including inpainting, avatar replacement, and generative content replacement \cite{li2017jpeg, hasan2017cartooning, xu2024examining, khamis2022deepfakes}. Various studies have examined the effectiveness of these methods across attributes \cite{hasan2018viewer, hasan2019can, li2017effectiveness}. Notably, Li et al. identified eight image obfuscation techniques from literature and studied the effectiveness of different protection methods across various attributes \cite{li2017effectiveness}. With the advent of image generation models, recent studies have proposed generative content replacement as an obfuscation technique and found them to be effective \cite{xu2024examining, khamis2022deepfakes}.

These studies provide a comprehensive understanding of image obfuscation techniques within the research community, though their practical application remains limited among end-users. Our tool addresses this gap by raising awareness of the various obfuscation techniques and their effectiveness. Existing tools, ranging from advanced editors to simplified redaction apps \cite{photoshop, obscuracam, dama2025}, either require significant expertise or focus narrowly on specific elements like faces or text, and in both cases, offer little support for user intent or contextual awareness, making them difficult for end-users to use effectively. To overcome this barrier, we enable easy application of image obfuscation techniques through our tool. Xu et al. \cite{xu2024examining} and Li et al. \cite{li2017effectiveness}, in their respective studies, emphasized the need for a human-in-the-loop image privacy protection tool. Responding to this call, we adopt a user-centered design approach, to develop an AI copilot, bridging the gap between advanced obfuscation techniques and end-user application.

\subsection{Automated Systems for Image Privacy Identification and Protection}
Researchers have made significant strides in developing automated tools for identifying and protecting sensitive content in images. Early efforts focused on specific elements, such as Hasan et al.'s tool to distinguish bystanders from subjects in photos \cite{hasan2020automatically} and Korayem et al.'s work on detecting screens \cite{korayem2016enhancing}. Expanding on this body of work, researchers also began integrating automatic obfuscation techniques. For instance, Ilya et al.'s Face/Off system recognizes and blurs faces for which the owner lacks permission \cite{ilia2015face}, while Frome et al. created a system for Google Street View that automatically detects and blurs faces and license plates \cite{frome2009large}. More recently, automated obfuscation techniques have also been extended to video and device-based privacy contexts to replace sensitive content in video streams \cite{hasan2017cartooning, iravantchi2024privacylens}.

As the field progressed, researchers also shifted towards more user-centered approaches. Li et al. proposed design considerations for an image obfuscation tool based on a Wizard of Oz study \cite{li2022obfuscation}. Additionally, Vishwamitra et al. introduced AutoPri, a novel system enabling automatic and user-specific content-based photo privacy control \cite{vishwamitra2022towards}. However, all automated obfuscation systems frame privacy as a classification task: identifying sensitive content and applying traditional obfuscation (e.g., blurring, masking) in a prescriptive, context-agnostic way. They offer limited user control and rarely support understanding why elements are risky or how risks relate to sharing intent.

Our work both extends and challenges this line of work by re-imagining privacy protection not just as a matter of automation accuracy, but as a process of helping users balance the privacy–publicity trade-off. We argue that users need to be active participants in this pipeline, as the boundary between privacy and publicity depends on contextual knowledge and intent. Our approach, realized through a formative study, moves beyond automation toward a copilot model, where the AI and user collaboratively manage image privacy.

\section{Formative Study}
We build on prior research demonstrating the potential of AI for image privacy \cite{xu2024examining, liu2022privacy}, by exploring how AI-powered features can simplify the process of identifying and mitigating image privacy risks, without requiring specialized expertise. To inform the design of our intelligent image obfuscation tool, we conducted a formative study with image editing experts. This study aimed to uncover user needs beyond basic usability barriers, focusing instead on their workflows, decision-making processes, the techniques they employ, and the challenges they encounter.

\subsection{Participants}
We recruited seven image manipulation experts (E1-E7) with 18 months to 10 years of experience in image editing, with some being self-taught (E3, E5) and others having taken design and digital art courses (E1, E2, E4, E6, E7).

\subsection{Study Procedure}
The study consisted of two major components (1) an image obfuscation task, and (2) a post-task semi-structured interview. The task was screen recorded and the interview was audio recorded. After obtaining informed consent, participants completed a brief demographic questionnaire and received an introduction to the concept of ``image obfuscation'' and the study's goals.

For the image obfuscation task, participants were asked to select images from a subset of the DIPA datasets \cite{xu2024dipa2, xu2023dipa}, which is a recent dataset that has records of images, annotated sensitive elements, the associated risks, and the sensitivity of the element \cite{xu2024dipa2, xu2023dipa}. The subset included over 115 images that were high quality and tagged as having sensitive content that had a user-reported score of more than 5 out of 7 to ensure that the images were ones that participants would have concerns about. Participants were asked to envision a relevant sharing intent and a privacy concern for each image they chose. Participants were then provided with Krita \cite{krita}, an open-source image editor which we additionally equipped with AI tools such as object segmentation, bounding box segmentation, text-based generation replacement, reference image-based generation, and avatar replacement. Participants were instructed to use these tools to obfuscate their chosen images for privacy preservation. To contextualize their work, we provided participants with privacy knowledge materials, including: 1) A list of potential sensitive content in images 2) Potential threats associated with image sharing 3) Examples of obfuscated images (before and after) from previous work \cite{xu2024examining, li2017effectiveness}, and a few examples the authors created. 

Finally, for the post-task semi-structured interview, we asked participants questions about the images they obfuscated, their workflow and thought process, the rationale behind their choices, their task experience, and suggestions and aspirations they have for an image obfuscation tool.

\subsection{Analysis}
To elicit user requirements for a system, we conducted a two-phased analysis.
First, we compared and contrasted the risks participants identified in the images they chose, to the risks that were pre-identified in the DIPA dataset \cite{xu2024dipa2,xu2023dipa}. Doing so allowed us to codify the elements and risks identified by our participants, which they identified beyond what was tagged in the original dataset, and which risks they seemed to miss.
Second, we began with open coding of the interview transcripts, followed by a thematic analysis to identify patterns across participants \cite{braun2006using}.
Two researchers coded the transcripts independently, and then later came together to resolve any conflicts in coding.
We also analyzed the screen recordings to understand how the users executed the tasks.

\subsection{Findings}
Our analysis revealed several findings, some of which confirmed previous knowledge, while others provided novel insights. Echoing prior literature \cite{xu2024examining}, participants (E2, E4, E6, E7) identified both advantages and drawbacks of using AI-powered image editing techniques for privacy. For example, participants appreciated that AI techniques allow for natural outputs and faster editing: ``I like the generative tool because it's really easy. You just click, and it's really good at creating, like, believable fake additions.'' However, our study revealed that users also appreciated the determinism and reliability of non-AI editing techniques. The stochastic nature of AI outputs occasionally frustrated participants:``it generated weird people .. it looks like witch-craft .. makes it nonsensical'' (E2). 

Also akin to prior work \cite{li2022obfuscation, vishwamitra2022towards}, we found that users followed a fairly standard workflow consisting of: Identifying sensitive content, selecting the sensitive content, and applying an obfuscation technique (E1-7). However, we also identified previously undocumented pain points with image editing for privacy. These challenges sharpened our understanding of the goal of obfuscation---to find a user-acceptable balance between reaping the social advantages of sharing personal information and mitigating privacy risks. Through our analysis of user needs, we also realized an underlying theme: while participants appreciated automation, they frequently also wanted control to express their preferences. This finding further pointed to collaborative, co-pilot style system rather than one that fully automates the process of identifying and mitigating privacy risks. We next discuss these pain points and design opportunities as they relate to three different phases of the standard workflow we identified.

\subsubsection{Identifying sensitive content --- Pain points and design opportunities}

\paragraph{Sharing intent and lived privacy concern impact what people want to obfuscate.}
In a few cases, identifying sensitive content was simple for our participants. Indeed, we found that participants consistently associated certain objects in images with privacy risks: e.g., license plates and laptop screens (E1, E2, E4, E6). ``I obscure the number plate...I think that's obvious''(E2). But there were other situations where this identification process was more nuanced. We found that when different participants chose the same image to obfuscate, they often chose \textit{different} sensitive objects within the image to obfuscate.

When exploring why, we found that participants were heavily influenced by their envisioned sharing intent and privacy concern (E1-7): ``maybe your family and friends aren't comfortable with the fact that you're gay'' (E1); ``maybe blonde hair is identifiable in this country'' (E4). This finding --- that different people have different intentions and concerns and thus take different approaches to mitigating privacy risks on the same image --- led us to realize there is \textbf{a need for a flexible, content-aware system that can take into account users' sharing intent and privacy concerns}.

\paragraph{Participants needed content-relevant reminders of privacy risk.}
During the study, some participants referred to the provided list of privacy risks to identify sensitive content in images. Several found it helpful when identifying threats (E1, E3, E6). ``it gave me a good idea of what different options I could do and what elements I should look for''. However, others engaged with it at a surface level, describing it as too large or detailed (E1, E2, E6, E7). They noted that while the list was helpful, it wasn't convenient: ``just knowing the list was helpful, but I would read it once and not really read it again'' (E2). Notably, some participants also said the list surfaced risks they had forgotten or hadn't previously considered (E1, E3, E4). ``I knew about some... but it looks like there is more that I did not know .. that even appearance and self-presentation could be privacy concerns'' (E3). Taken together, these findings suggest that \textbf{users need to be made aware of the broad spectrum of content-specific privacy risks}.

\subsubsection{Selecting Sensitive Content --- Pain points and design opportunities}
Despite all participants having more than one year of experience with image editing, they struggled with the process of selecting sensitive content in an image after they had identified that content as sensitive. For example, they faced difficulties in precisely selecting an object or ensuring that the correct layer was selected when manipulating the image (E1, E4-7).``it was a little confusing for me with all the layer stuff, but it wasn't super confusing, especially since I had prior experience'' (E1). Participants preferred to use the object selection tool (an AI-powered object segmentation tool) to precisely select an area of interest (E1, E2, E4). This observation further highlighted the \textbf{need to reduce the complexity in directly selecting and manipulating sensitive content in images}.

\subsubsection{Obfuscating Images --- Pain points and design opportunities}

\paragraph{Participants select obfuscation techniques based on prior familiarity, not effectiveness at mitigating privacy risks.}
Some participants implicitly considered the pros and cons of different obfuscation techniques. For example, one participant chose image generation over blurring as it produced ``believable fake additions'' (E4). However, many participants selected an obfuscation technique based on convenience and prior familiarity, rather than based on the appropriateness of a technique vis-a-vis a specific privacy risk and/or sharing intent (E1, E2, E4-6). For instance, participants often selected pixelation or blurring, even though these techniques were less effective than other options. When asked why they chose to use the same obfuscation techniques and not consider others, E6 rationalized:  ``I just forgot how to do it.''
These observations highlight the \textbf{need for a system that makes users aware of and understand the pros and cons of broad spectrum of obfuscation techniques}.

\paragraph{The most appropriate obfuscation technique varied based on sharing intent and participant preference.}
Our analysis showed that personal preference and sharing intent were the most important factors for participants when choosing obfuscation techniques.
For instance, E2 chose to blur a bystander in an image for a research paper because they wanted the obfuscation to be apparent to the viewer.

Others experimented with different obfuscation techniques and ultimately selected one that best matched their preference and comfort level.
We noted that participants exhibited varying levels of comfort with sharing obfuscated images---E2 was reluctant to share a heavily obfuscated image online, while E7 was more open, even after heavily manipulating an image by adding obvious fictional elements, like a snow castle in the background.
Participants also aimed to more precisely control the output of AI-powered obfuscation techniques by using prompts and reference images. E3, for example, replaced the face of a bystander in an image with one of a specific public figure by using a reference image of that public figure: ``I replaced the face .. with the photo of a model... models are like very much public'' (E3).
Overall, these findings further highlight that a purely automated approach to obfuscation is unlikely to be widely accepted: \textbf{users want granular control when applying and fine-tuning obfuscations}.

\section{Design Requirements for an AI-Powered Image Privacy Copilot}

To summarize, from our formative study, we distilled five design requirements for an image privacy copilot that helps users navigate the privacy / publicity boundary when sharing images online.

\begin{enumerate}
    \item[\textbf{DR1}] \textbf{Enable expressive articulation of privacy concern and sharing intent}. Participants who selected the same image to obfuscate in our formative study often flagged different regions of the image as sensitive. Their selections, in turn, were driven in large part by \textit{who} they were concerned about sharing the image with and \textit{why} they were sharing the image in the first place. Thus, an image privacy copilot should adapt to users' directly expressed privacy concerns and sharing intent.
    \item[\textbf{DR2}] \textbf{Increase awareness of content-level privacy risks}. Our participants found external memory aids of potential privacy risks to be helpful when identifying risky content in images. Many people may be unaware of some types of privacy risks, for example. Accordingly, an image privacy copilot should also proactively surface potentially risky content beyond a user's immediate privacy concern.
    
    \item[\textbf{DR3}] \textbf{Promote informed decision-making}. Ultimately, the goal of an image privacy copilot should be to maximally mitigate privacy risk while minimally compromising the user's sharing intent. Sometimes, sharing risky content is necessary to communicate intent --- some participants even lamented the idea of sharing artificial images. To promote informed decision-making, an image privacy copilot should provide explanations to users of privacy risks it identifies, and promote obfuscation strategies that minimally interfere with sharing intent.

    \item[\textbf{DR4}] \textbf{Facilitate easy and effective application of obfuscation techniques}. Even our expert image-editing participants struggled with applying all but the most basic obfuscation strategies (e.g., blurring). Many selected obfuscation techniques are based strictly on prior familiarity and not effectiveness, rarely considering specific privacy concerns nor sharing intent. Accordingly, an image privacy copilot should present users with a streamlined process to reduce the complexities of precisely selecting the sensitive content and effectively applying the obfuscation techniques.
    
    \item[\textbf{DR5}] \textbf{Ensure autonomy and granular control}. Ultimately, our participants cared not just about mitigating privacy risks with obfuscation techniques, but also about preserving image authenticity and aesthetics. Moreover, their preferences for what and how to obfuscate varied based on their privacy concerns, sharing intent, and personal preferences. Accordingly, an image privacy copilot should remain just that --- a copilot. It should afford users ultimate authority to make decisions about what and how to obfuscate an image.
\end{enumerate}

\begin{figure*}[ht]
    \centering
    \includegraphics[width=\linewidth]{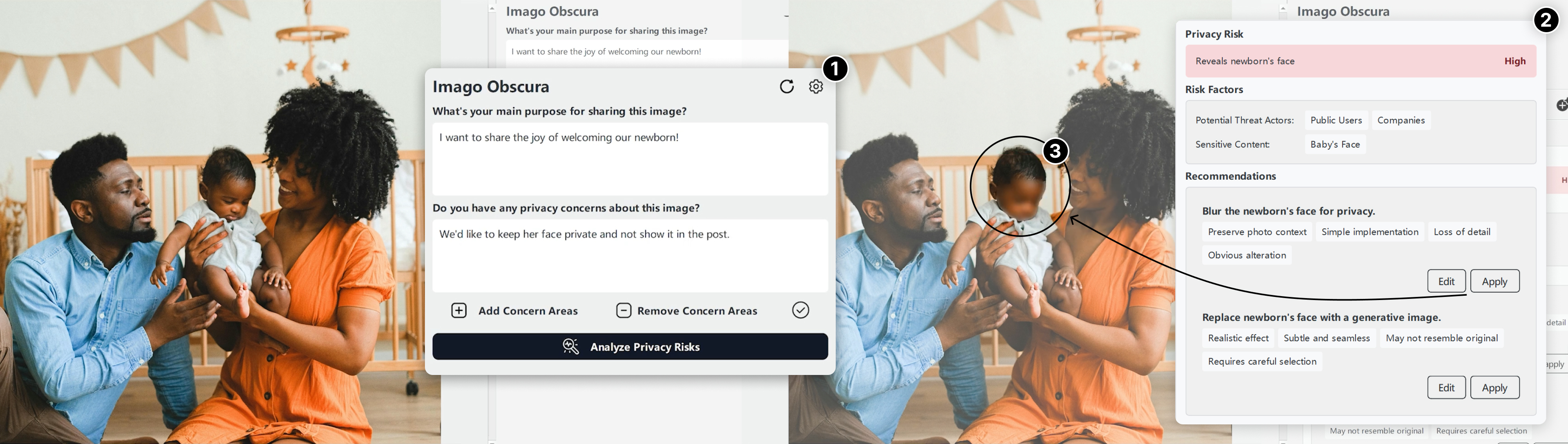}
    \vspace{-20pt}
    \caption{\systemname{} enables users to express their sharing intent and their privacy concerns in natural language, subsequently identifying pertinent risks and recommending obfuscation techniques.}
    \Description{A family photo is shown. The user specifies a desire to share the image but not show the newborn’s face. The system flags the baby’s face as high-risk and recommends blurring or generative replacement. The final image has the baby’s face blurred.}
    \label{fig:text-articulation}
\end{figure*}

\begin{figure*}[ht]
    \centering
    \includegraphics[width=\linewidth]{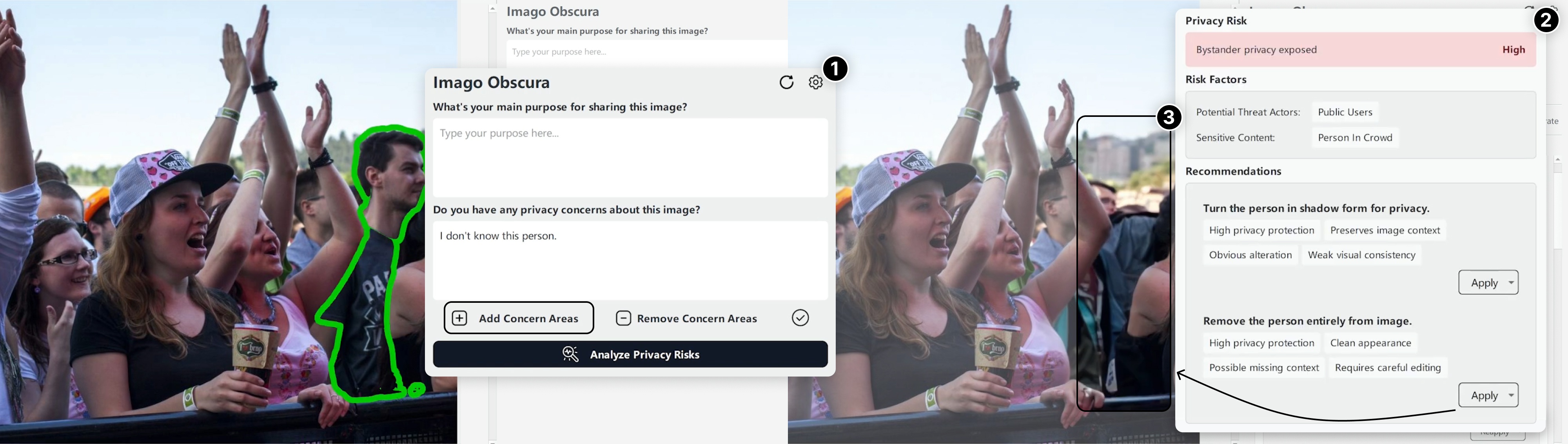}
    \vspace{-20pt}
    \caption{\systemname{} enables the user to directly select areas of concerns which the tool will automatically precisely select and highlight in green, subsequently identifying pertinent risks and recommending obfuscation techniques.}
    \Description{The user clicks on an unfamiliar person in a crowd photo. The system highlights the contour of the selected individual, the system then identifies the bystander as a privacy risk and suggests replacing them with a generative image or converting them into a silhouette. The final image applies the selected obfuscation.}
    \label{fig:visual-annotation}
\end{figure*}

\section{\systemname{}}

Guided by the design requirements we distilled from our formative study, \systemname{} enables users to make informed decisions about if and how to obfuscate personal images by guiding users through a structured workflow. Throughout this section, we provide examples of how we operationalized the design requirements in \systemname. Figure~\ref{fig:teaser} shows the entire user workflow in sequence.

\subsection{\textbf{DR1: Enabling expressive articulation of privacy concern and sharing intent}}

\paragraph{Natural language expression.} \systemname{} allows users to directly articulate their privacy concerns and sharing intent through natural language. To scaffold this articulation, we ask users two questions inspired by a study conducted in previous work \cite{li2022obfuscation} --- i.e., ``What's your main purpose of sharing this image?'' and ``Do you have any privacy concerns about this image?''. For example, in Figure \ref{fig:text-articulation} the user expresses that they want to announce the birth of their child, but also do not want to show her face.

\paragraph{Visual annotation of areas of concern.} Natural language is powerful, but sometimes it is faster and quicker for users to directly select and manipulate sensitive content in images. For example, in Figure \ref{fig:visual-annotation}, there are multiple people in the photo: it would be easier to directly click on the bystander's face than to type out a description of the bystander. Therefore, we also gave users the option to directly select content in an image with which they were concerned.

\subsection{\textbf{DR2: Increasing awareness of content-level privacy risks}}
Identifying and addressing privacy risks associated with image sharing is essential for informed decision-making.
\systemname{} identifies five content-related categories of image privacy risk, drawn from an analysis of prior art that includes taxonomies of sensitive content \cite{li2020towards, li2018human, xu2024dipa2} and image privacy risk at large \label{sec:risks-lit}\cite{ahern2007over, xu2024dipa2, fire2014online, patsakis2015privacy}. 

Below, we describe each of the five image privacy risks \systemname{} identifies, and provide illustrative examples of each.

\begin{figure*}[htb]
    \centering
    \includegraphics[width=\linewidth]{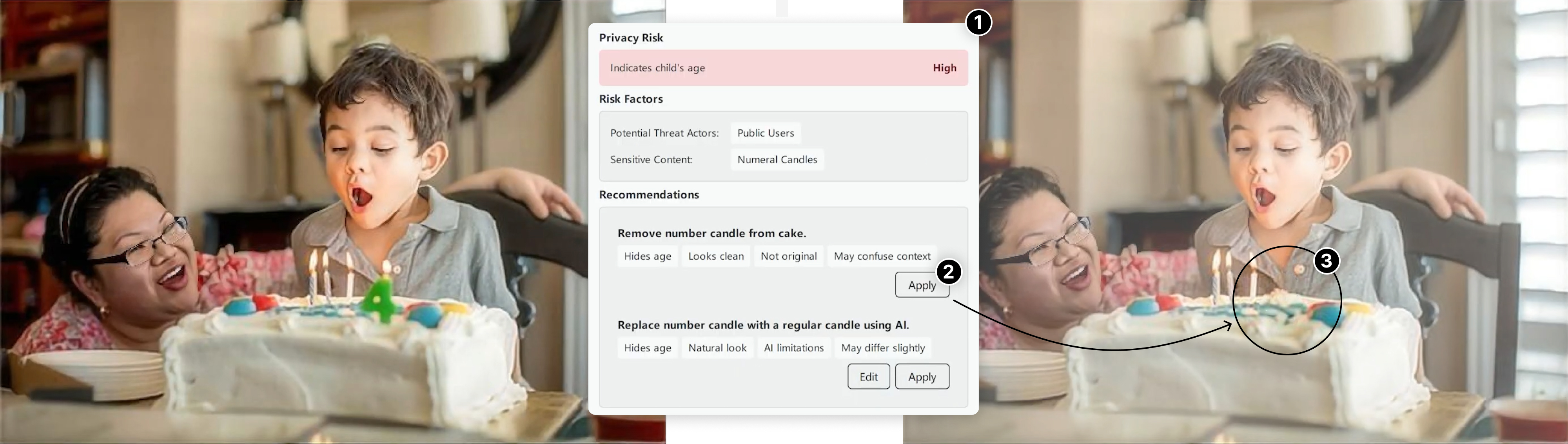}
    \vspace{-20pt}
    \caption{\systemname{} addresses ``self disclosure risks''. (1) Identifies that the numbered candle can reveal personal information. (2) Recommends removing the candle from the image. (3) Precisely selects the sensitive area, the candle, and applies inpainting.}
    \Description{A three-panel image demonstrating Imago Obscura addressing "self disclosure risks". The left panel shows a child blowing out candles on a birthday cake. The center panel displays the tool's interface, identifying "Indicates child’s age" as a high privacy risk and recommending to remove the number candle from the cake. The right panel shows the same image with the candle digitally removed.}
    \label{fig:self-disclosure}
\end{figure*}

\begin{figure*}[htb]
    \centering
    \includegraphics[width=\linewidth]{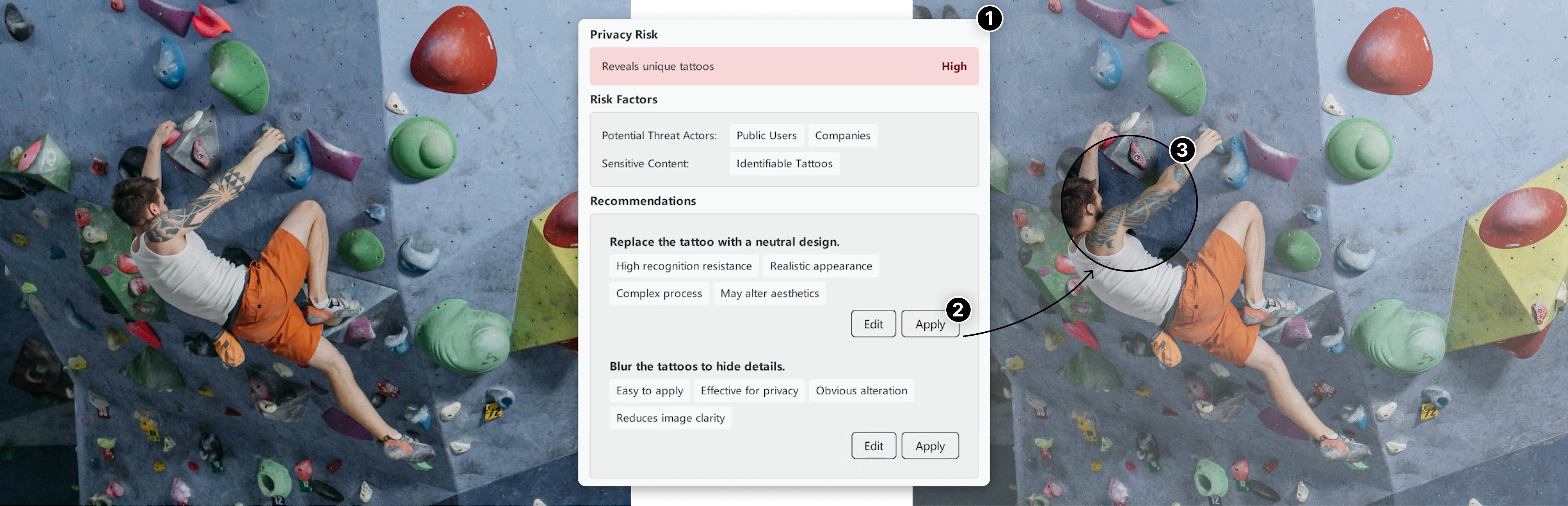}
    \vspace{-20pt}
    \caption{\systemname{} addresses ``identity exposure risk''. (1) Identifies that the tattoo can reveal the person's identity. (2) Recommends to replace the tattoo with a new one. (3) Precisely selects the sensitive area, the tattoo, and applies generative content replacement.}
    \Description{A three-panel image illustrating Imago Obscura addressing "identity exposure risk". The left panel shows a person climbing with a visible arm tattoo. The center panel displays the tool's interface, identifying "Reveals unique tattoos" as a high privacy risk and recommending to replace the tattoo with a neutral design. The right panel shows the same image with the tattoo replaced by generative content.}
    \label{fig:identity}
\end{figure*}

\paragraph{Self-Disclosure Risk}
Self-disclosure risk involves the unintended revelation of personal details that may compromise privacy. This risk occurs when subtle cues in an image expose personal habits, health conditions, life events, etc, or allow outsiders to infer details about someone's interests, affiliations, or habits: e.g., a photo of a bookshelf might suggest certain intellectual or political leanings; a gym bag could hint at fitness routines; medication bottles can disclose health issues; specific types of food may imply dietary restrictions or choices. In Figure \ref{fig:self-disclosure}, a mother wanting to share a photo of her child cutting a birthday cake may unwittingly reveal the child's age through a numbered candle. \systemname{} identifies this detail and flags it as a self-disclosure risk.

\paragraph{Identity Exposure Risk}
Identity exposure risk refers to the potential for an individual's identity to be uncovered through visible personally identifiable information such as facial features, ID cards, or mail. As shown in Figure \ref{fig:identity}, a climbing gym photo intended for the website's hero image includes a person whose face is obscured, but a visible tattoo could still reveal their identity.

\begin{figure*}[htb]
    \centering
    \includegraphics[width=\linewidth]{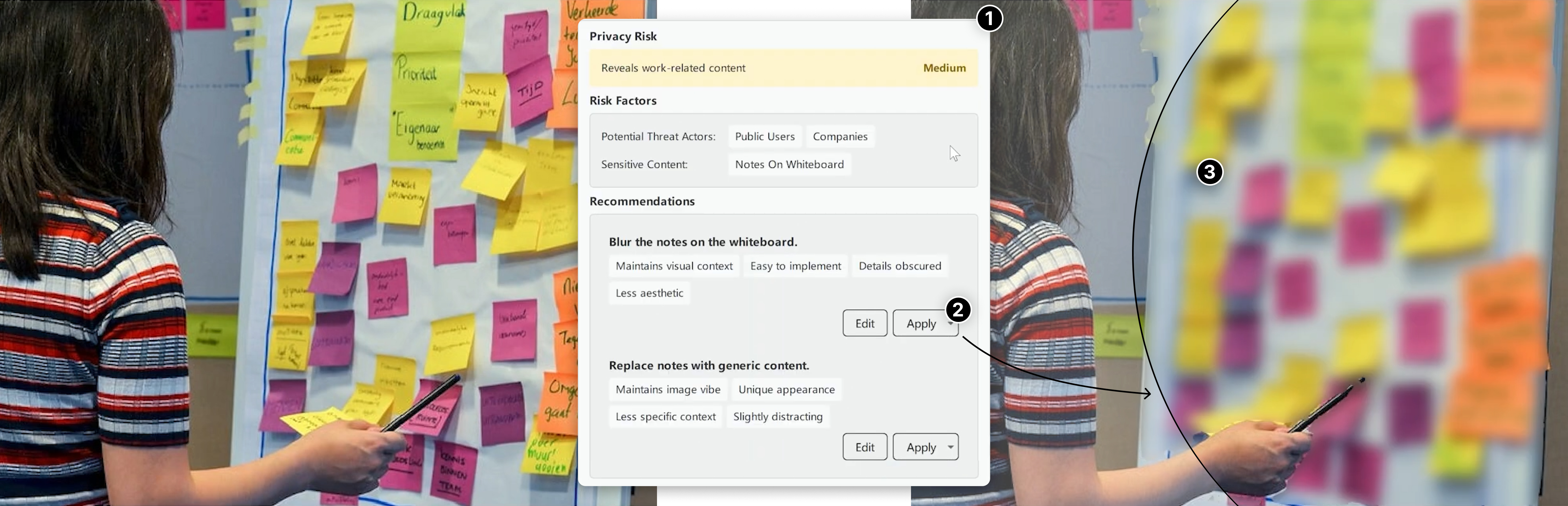}
    \vspace{-20pt}
    \caption{\systemname{} addresses ``confidential information leakage risk''. (1) Identifies that the notes on the board can reveal confidential information. (2) Recommends to blur the notes on the board. (3) Precisely selects the sensitive area, the board, and applies blur.}
    \Description{A three-panel image demonstrating Imago Obscura addressing "confidential information leakage risk". The left panel shows a person standing in front of a whiteboard covered in colorful sticky notes. The center panel displays the tool's interface, identifying "Reveals work-related content" as a medium privacy risk and recommending to blur the notes on the whiteboard. The right panel shows the same image with the whiteboard and sticky notes blurred.}
    \label{fig:confidential}
\end{figure*}

\begin{figure*}[htb]
    \centering
    \includegraphics[width=\linewidth]{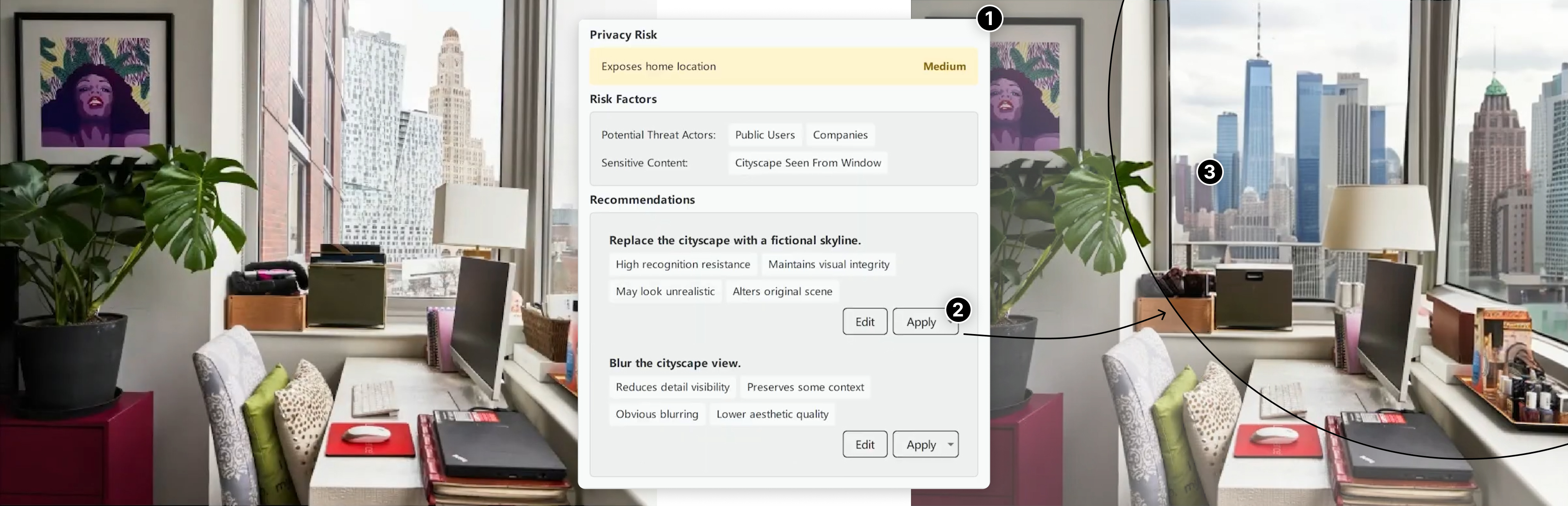}
    \vspace{-20pt}
    \caption{\systemname{} addresses ``location exposure risk''. (1) Identifies that the window view can reveal the location. (2) Recommends to replace the window view. (3) Precisely selects the sensitive area, the window, and applies generative content replacement.}
    \Description{A three-panel image showing Imago Obscura addressing "location exposure risk". The left panel displays a home workspace with a window view revealing a recognizable city skyline. The center panel shows the tool's interface, identifying "Exposes home location" as a medium privacy risk and recommending to replace the cityscape with a fictional skyline. The right panel shows the same image with the city view digitally replaced.}
    \label{fig:location}
\end{figure*}

\paragraph{Confidential Information Leakage Risk}
Confidential information leakage risks occur when secret or proprietary information is visible in the background of an image. This can happen, for example, in business environments where whiteboards or computer screens are captured or documents are visible on a desk. For instance, in Figure \ref{fig:confidential}, a researcher planning to share a photo from a collaborative workshop captures notes on a whiteboard. \systemname{} flags these notes as having the potential for confidential information leakage and suggests techniques to mitigate this risk.

\paragraph{Location Exposure Risk}
Location exposure risk involves identifiable location information being revealed in an image, compromising the user's physical privacy and safety. These risks might occur when architectural features, recognizable landmarks, or specific weather patterns are visible. For instance, in Figure \ref{fig:location}, a person wants to share a picture of their home office setup. The view from the window shows a few recognizable buildings, potentially disclosing the exact location of their home. \systemname{} identifies these details and alerts the user to potential location exposure risks.

\begin{figure*}[htb]
    \centering
    \includegraphics[width=\linewidth]{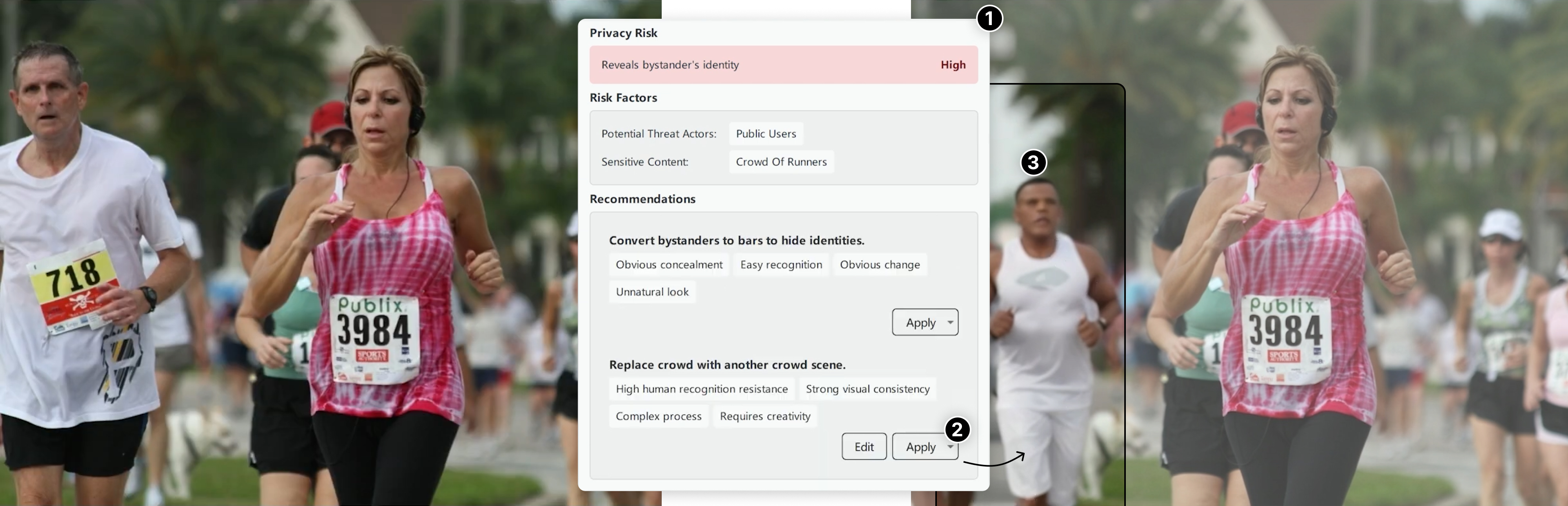}
    \vspace{-20pt}
    \caption{\systemname{} addresses ``bystander privacy risk''. (1) Identifies that the bystanders' privacy might be at risk. (2) Recommends to generate a new running crowd scene. (3) Precisely selects the sensitive area, the bystander, and applies generative content replacement.}
    \Description{A three-panel image illustrating Imago Obscura addressing "bystander privacy risk". The left panel shows a marathon scene with a runner, in the corner is a participant who was not intended to be captured. The center panel displays the tool's interface, identifying "Reveals bystander’s identity" as a high privacy risk and recommending to generate a new crowd scene. The right panel shows the same image with the fellow participant in the background replaced by generative content.}
    \label{fig:bystander}
\end{figure*}

\paragraph{Bystander Risk}
Bystander risks arise when individuals in the background of an image are unintentionally captured. This can occur in crowded public places or events, where bystanders may not be aware that they are being photographed. Examples include street scenes, public gatherings, or casual photos taken in parks. In Figure \ref{fig:bystander}, a marathon runner shares a picture of themselves running, but a bystander's face is visible in the background. \systemname{} flags the bystander, suggesting the use of techniques to obscure their identity.

\subsection{\textbf{DR3: Promote informed decision-making}}
To promote informed decision-making, we provide users with detailed explanations of both the risks identified and obfuscation techniques recommended to address those risks.

\paragraph{Presenting Risks, Sensitive Content, Threat Actors, and Severity} While surfacing pertinent risks is crucial, we realized it is equally important to present them in a way that is easy to understand and act upon. To ensure user comprehension, we describe risks in natural language, aligned with the user’s concerns and sharing intent. For example, in Figure \ref{fig:location}, \systemname{} flags a location exposure risk with the label ``Exposes your location'' and provides an explanation of of the source of the risk: ``Cityscape seen from window''. We also present the sensitive content from which the risk arises and potential threat actors who might be able to exploit the risk. Recognizing that some sensitive elements can reveal more than others and that risks vary in severity, we also classify and present the severity of each risk as High, Medium, or Low.

\paragraph{Presenting Image Obfuscation Techniques and Their Attributes}
Our formative study revealed that participants often chose image obfuscation techniques arbitrarily, partly due to forgetting about the gamut of available options.  Therefore, beyond risk identification, \systemname{} presents obfuscation techniques for each identified risk: converting the recall problem into one of recognition. Our tool enables the application of a broad range of obfuscation techniques curated from existing literature \cite{li2017effectiveness, xu2024examining, khamis2022deepfakes}. The list of curated obfuscation techniques is in Appendix \ref{sec:obf-lit}, and also visualized in Figure~\ref{fig:obfuscation}.
The formative study also showed that participants often overlooked the effect of obfuscation techniques on the final image. Accordingly, we highlight each technique's unique properties and how it affects the image. Informed by prior art, we consider various attributes to define the effectiveness of each obfuscation technique: effectiveness against recognition, detectability, visual harmony, narrative coherence, realism, and vulnerability \cite{li2017effectiveness, hasan2018viewer, xu2024examining}. These attributes are presented in Appendix Table \ref{tab:technique_attributes}.

\subsection{\textbf{DR4: Facilitate easy and effective application of obfuscation techniques}}
To simplify the mitigation privacy risks that users want to address, we made both selection of risky content pertinent to those risks and application of obfuscation techniques accessible through one-click interactions.

\paragraph{Precise selection of risky content.} After a user chooses which risk they would like to address, \systemname{} enables the selection of the sensitive content pertinent to the risk automatically with a one-click action. The system then precisely selects the sensitive content and awaits confirmation from the user.

\begin{figure*}
    \centering
    \includegraphics[width=0.49\textwidth]{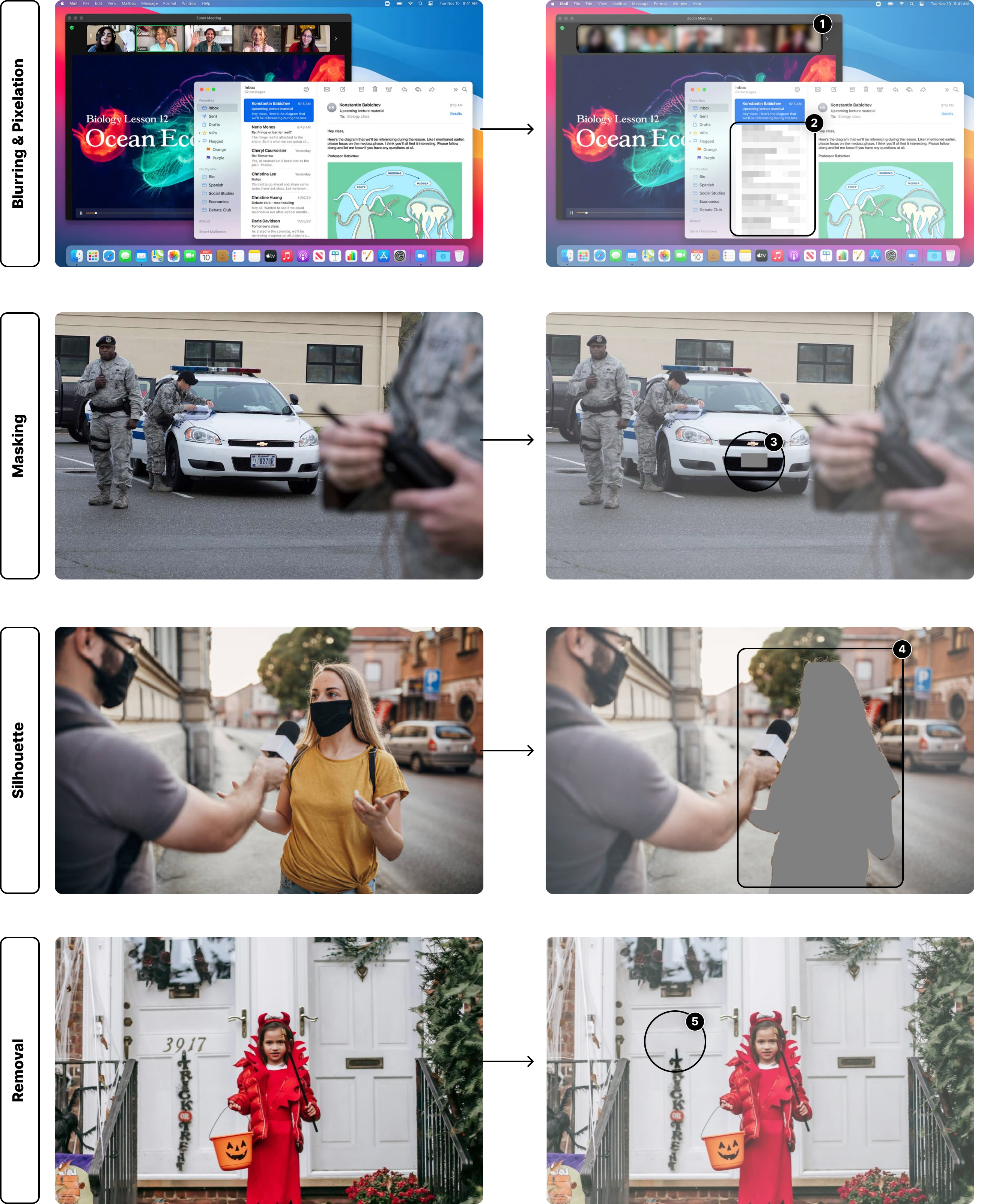}
    \includegraphics[width=0.49\textwidth]{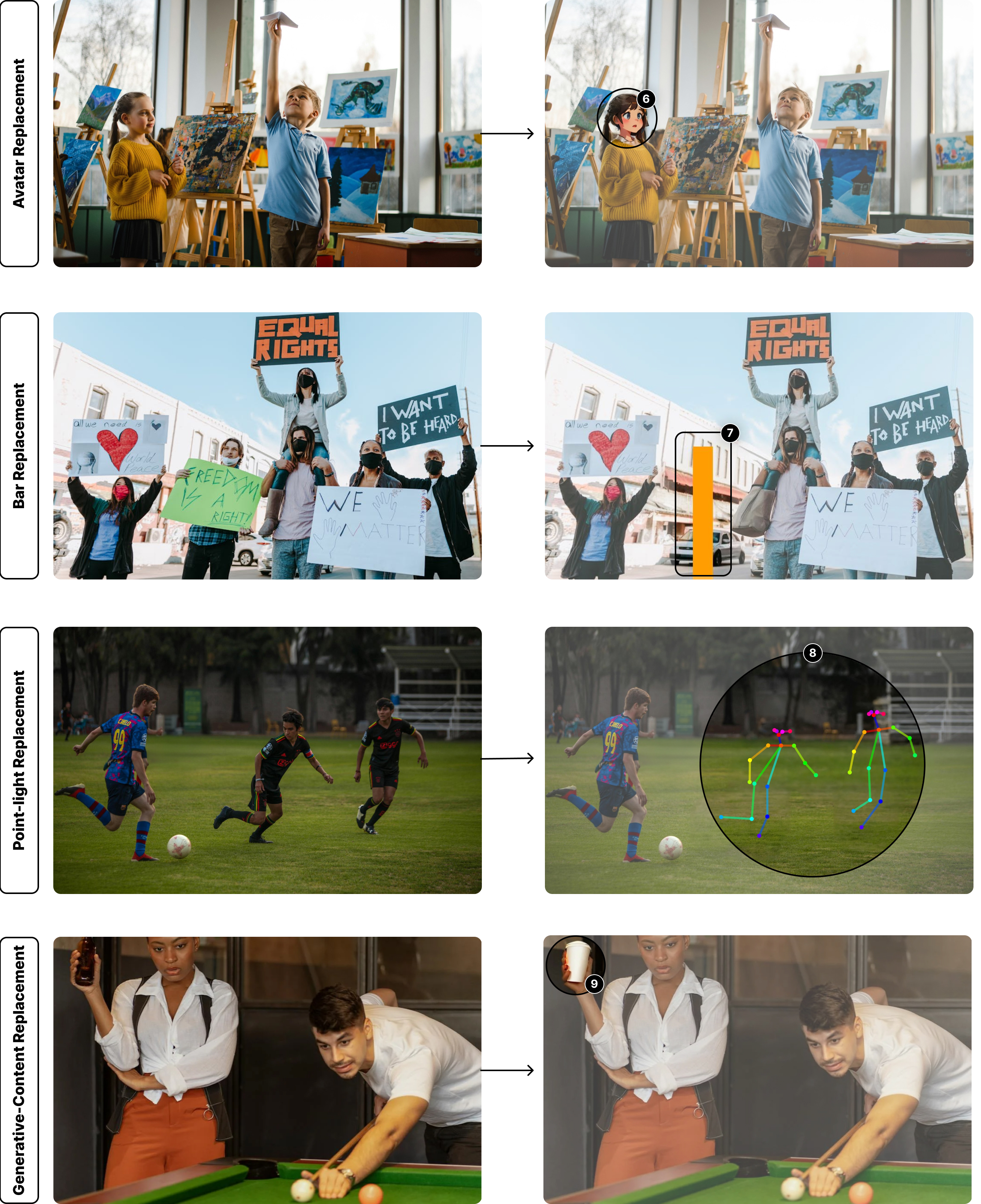}
    % \vspace{-5pt}
    \caption{Demonstration of the diverse image obfuscation techniques enabled by \systemname. Each pair shows the original (left) and obfuscated (right) image: (1-2) \textit{Blurring} and \textit{pixelation} of screen content to protect confidential information (3) \textit{Masking} of license plate to preserve vehicle anonymity (4) \textit{Silhouette masking} to anonymize a whistle-blower in a news article (5) \textit{Removal} of house number to conceal the specific location (6) \textit{Avatar replacement} to protect identity of a child's friend (7) \textit{Bar replacement} to obscure a fellow participant (8) \textit{Point-light representation} shows body pose while preserving anonymity; (9) \textit{Generative replacement} of an alcohol bottle avoids promoting alcohol.}
    \Description{Each pair shows the original (left) and obfuscated (right) image:
    (1–2) A laptop screen with sensitive content is obfuscated using blurring and pixelation to protect confidential information.
    (3) A car's license plate is masked to preserve vehicle anonymity.
    (4) A whistle-blower speaking for a news channel is anonymized using silhouette masking.
    (5) A child in a halloween costume is pictured at a doorstep; the house number in the background is removed to conceal the specific location.
    (6) A classroom photo where a child’s friend is replaced with an avatar to protect their identity.
    (7) A group protest photo where the only unmasked participant is obscured with a bar replacement to prevent identification.
    (8) A soccer scene where 2 of 3 players are represented with point-light figures to preserve pose while hiding identity.
    (9) A person playing pool next to a person holding a alcohol bottle, which is replaced with a disposable glass with generative content to avoid promoting alcohol.}
    \label{fig:obfuscation}
\end{figure*}

\paragraph{Easy application of obfuscation techniques.} On confirmation that the selection aligns with the user's intention, \systemname{} automatically applies the chosen obfuscation technique, also with a one-click interaction. Figure \ref{fig:obfuscation} illustrates the different obfuscation techniques that are possible through \systemname{} --- these include traditional image transformation techniques like blurring, pixelation, masking, silhouette masking, as well as AI-powered techniques like removal/inpainting, bar replacement, point light replacement, avatar replacement, and generative content replacement.

\subsection{\textbf{DR5: Ensure autonomy and granular control}}
To ensure autonomy and granular control, \systemname{} affords users choice at every step of the risk identification and mitigation workflow. 

\paragraph{Choice over which risks to mitigate, and how to mitigate them} 
While we present various risks pertinent to the image and the user's sharing intent, users retain full control over which privacy risks they want to address. For instance, if a user is sharing a photo of themselves in front of the Eiffel Tower, they can make an intentional choice to forgo location exposure risks. For each sensitive content segment linked to a risk, the system recommends at least two obfuscation techniques from which the user can choose.

\paragraph{Manual refinement of automatic selection}
While the system automatically selects sensitive objects for users to obfuscate to address a specific risk, they are also afforded the option to refine their selection of objects in the image prior to applying the recommended obfuscation technique.

\paragraph{Ad hoc use of obfuscation techniques}
Beyond the recommended obfuscation techniques, \systemname{} also enables users to apply obfuscation techniques in an ad hoc manner.
It does so by presenting a toolbar with an AI-powered precise selection option, two traditional image transformations (blurring and masking), and two AI-powered image generation-based obfuscation techniques (generative content replacement and avatar replacement).

\paragraph{Granular control over obfuscation techniques}
\systemname{} provides users with granular control over obfuscation techniques through the use of intensity control sliders, text prompts, and reference image upload options. 

For example, users can increase the blur on confidential information, replace a bystander's face with that of a reference photo, or create a fictional background for added narrative effect.

\section{Implementation}

\begin{figure*}[ht]
    \centering
    \includegraphics[width=\linewidth]{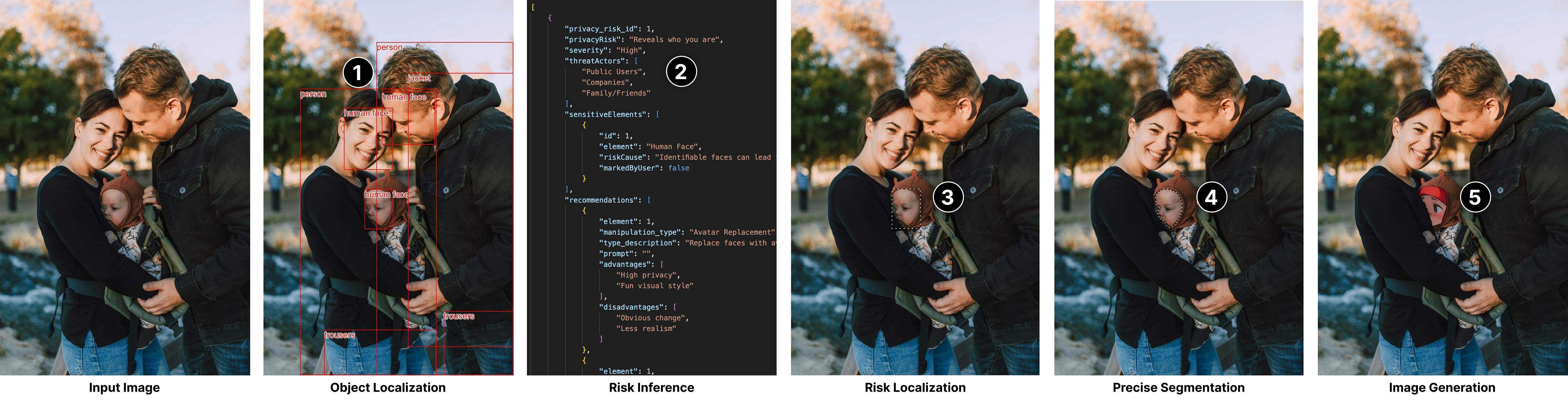}
    \vspace{-20pt}
    \caption{Step-by-step outputs of each model in the \systemname{} pipeline.  
    1) The vision model detects and labels objects with bounding boxes.  
    2) The MLLM identifies sensitive content and recommends obfuscation strategies (shown as a JSON object).  
    3) The vision model re-localizes the sensitive elements identified by the MLLM.  
    4) The segmentation model refines the selected region with precision.  
    5) The image generator replaces the selected region using the chosen obfuscation method.}
    \Description{
    A six-panel image showing the sequential steps of the Imago Obscura pipeline.
    The first panel shows the original family photo with a baby.
    The second panel displays labeled bounding boxes around detected objects such as person, jacket, human face, trousers, etc.
    The third panel shows a JSON-style risk inference output identifying faces as sensitive content and recommending avatar replacement.
    The fourth panel shows a box around the baby's face, indicating the sensitive objects location.
    The fourth panel shows the system selects the baby's face using a dotted outline around the contour, indicating precise region selection.
    The fifth panel shows the final obfuscated image with the baby's face replaced by a generative avatar while the rest of the image remains unchanged.
    }
    \label{fig:implementation}
\end{figure*}

We implemented \systemname{} as a plugin for the open-source graphics editor Krita \cite{krita}. \systemname{} utilizes an integrated ensemble of four AI models to enable the workflow and design space we described.
Specifically, it leverages a multimodal large language model, GPT-4o \cite{openai2023gpt, radford2021learning}, 
to identify privacy risks in images based on user's expressed privacy concerns and the taxonomy of privacy risks we described above;
a vision model, Florence 2 \cite{xiao2023florence2advancingunifiedrepresentation}, to automatically annotate images with bounding boxes and labels for objects found in the image; 
a segmentation model, SAM \cite{kirillov2023segment}, to get precise selections of sensitive content in the image, helping us to associate privacy concerns with specific regions of the image; and, a text-to-image generation model, stable diffusion \cite{karras2020training, rombach2022high}, to automatically apply AI-powered image obfuscation techniques if the user so chooses. A full overview of this process can be seen in Figure~\ref{fig:implementation}.

Users select an image, and can articulate their privacy concerns and sharing intent of that image through natural language and visual annotation. Once the user presses a button to analyze privacy risks, we feed the image, users' concerns and sharing intent, and the taxonomy of image privacy risks we synthesized from prior literature through our ensemble of multimodal AI models.
Using this input, \systemname{} identifies sensitive content in the image that users may consider obfuscating. These risks are presented to users in the form of explanations of why that content may be risky. Finally, users can choose to act on any of the identified risks. For each risk, the system presents a subset of relevant obfuscation techniques from the nine techniques we found in prior literature (Appendix \ref{subsec:obfuscation techniques}). Users can easily apply these techniques through simple click-based interactions.
The cumulative effect of this workflow is that users get highly customized and personalized assistance with identifying and mitigating pertinent privacy risks in images they hope to share online.

\subsection{Pre-scan Process}
Prior to engaging the Multi-modal Large Language Model (MLLM), the image undergoes a preliminary scan using the Florence vision model \cite{xiao2023florence2advancingunifiedrepresentation}, chosen for its robust object detection and classification capabilities. Florence provides labeled bounding boxes for all detected objects, which are then supplied to the MLLM. This preliminary analysis serves as a form of visual prompting, employing the ``set-of-mark prompting'' technique that has been shown to improve MLLM's visual reasoning capabilities \cite{yang2023set}. The annotated image functions as a visual guide, informing subsequent steps in the privacy risk analysis and obfuscation process, enabling the MLLM to focus on targeted areas within the image.

\subsection{Prompting Techniques to Identify Risks}
The tool begins by developing an understanding of the image, integrating the user's sharing intent and privacy concerns. This process is guided through a series of structured prompts (see Appendix \ref{sec:prompt-id}), using chain-of-thought prompting \cite{wei2022chain} to instruct the Multimodal Large Language Model (MLLM) to systematically analyze the image content, user context, and potential privacy risks.

The system initiates the analysis by instructing the MLLM to examine the image and user's sharing intent and privacy concerns highlighted through text or visual annotations. User-provided visual annotations, areas marked in green, are provided as visual prompts and guided to be interpreted as direct indicators of privacy concerns. The MLLM is specifically instructed to prioritize these user-indicated regions to ensure that the user's specific privacy concerns are addressed first. Once these concerns are identified, the MLLM follows a two-step process.
\begin{enumerate}
    \item \textit{Identify Sensitive Elements}: The model is guided to scan the image to identify potentially sensitive content by referring to a curated list of potential sensitive elements. This curated list, derived from prior literature, encompassed detailed elements in categories such as identity and personal information, nudity, and social contexts. 
    \item \textit{Assess Privacy Risks}: The MLLM is further guided to evaluate the identified sensitive elements to determine potential privacy risks, while referencing a curated list of privacy risks (Section \ref{sec:risks-lit}). These risks are assigned a severity level (High, Medium, Low) and potential threat actors (e.g., public users, companies, acquaintances) based on context and are instructed to be presented in natural language.
\end{enumerate}

\subsection{Prompting Technique to Recommend Obfuscation}
Following the identification of privacy risks, the system instructs the MLLM to identify and recommend appropriate image obfuscation techniques (see Appendix \ref{sec:prompt-rec} for the prompt). The model is prompted to choose from a list of obfuscation techniques (Appendix \ref{sec:obf-lit}). The MLLM is then instructed to generate up to two recommendations per sensitive element, using user-friendly, non-technical language to describe the obfuscation methods. These recommendations are tailored based on the image, the user's provided concern, and attributes identified in previous literature (Appendix Table \ref{tab:technique_attributes}). 
Finally,  we receive a JSON object that includes the risks, their severity, relevant threat actors, sensitive elements, and recommended obfuscation techniques with their attributes.

\subsection{Locating and Selecting Sensitive Elements}
To apply obfuscations precisely, the tool first aggregates all identified sensitive elements from the JSON object output by the MLLM. It then uses the Florence vision model to locate each sensitive element, generating bounding boxes to provide approximate locations. For more precision, these bounding boxes are passed to the Segment Anything Model (SAM) \cite{kirillov2023segment}, which generates detailed contours. Finally, the detailed contours are used by the graphic editor's selection tool to enable precise selection of the sensitive content.

\subsection{Applying Obfuscations}
\systemname{} applies traditional image obfuscations, such as blurring and pixelation, through the integrated image editing tools of Krita. For the AI-driven techniques, the system combines AI methods with these traditional transformations. Techniques like removal/inpainting, avatar replacement, and generative content replacement leverage a stable diffusion model to replace sensitive content with generated elements. For bar and point-light replacements, the sensitive element is first removed via inpainting, followed by the application of the respective replacement.

\section{Evaluation}
To evaluate \systemname{}, we conducted an in-person lab user study with 15 participants. The goal of our user study was three-fold. First, drawing on the Security and Privacy Acceptance Framework (SPAF), which outlines three key barriers that inhibit end-user adoption of new and expert-recommended security and privacy tools \cite{das2022security}, we wanted to assess \systemname's impact on users' awareness of, motivation to address, and their ability to mitigate pertinent privacy risks in images.
Next, we aimed to assess how well \systemname{} fulfilled the five design requirements we distilled from our formative study (DR1–5). Finally, we wanted to understand to what extent users found \systemname{} useful and usable. To these ends, we used pre-task and post-task surveys, a final survey including the System Usability Scale (SUS), and a semi-structured exit interview. This protocol was revised based on insights we gained from an initial set of pilot studies we conducted with a separate set of 12 participants.
Note that we also conducted a technical evaluation of the risk identification component of our model pipeline to ensure that its outputs were robust and accurate --- we share the details of that evaluation in the Appendix \ref{sec:tech-eval}. In short, with GPT-4o \cite{openai2023gpt}, our approach achieved an accuracy of $\sim$70\% for sensitive object identification, $\sim$83\% for risk category classification and $\sim$73\% for severity assessment on the DIPA2 dataset \cite{xu2024dipa2}. 
We consider this evaluation peripheral because, to some degree, \systemname{} is model agnostic --- if more accurate models become available in the future, \systemname{} will be able to take advantage of them. Moreover, since \systemname{} is a copilot and not a full automation tool, we consider this performance good enough to support users in making informed decisions. 

\subsection{Participants}
We recruited 15 end-users (P1-P15) who had previously shared personal images online. Participants ranged in age from 18 to 31 years old (5 male and 10 female). All participants had an academic background including undergraduate students, PhD candidates, and research assistants. Six participants reported previous experience with image obfuscation techniques, using tools such as Adobe Photoshop, Background Remover app, Adobe Firefly, Canva's blurring tool, and built-in smartphone editing features.

\subsection{Study Procedure}
Our study lasted approximately one hour. Participants were asked to bring four personal images each to the study: two they previously shared online (shared images) and two they wanted to share but had withheld due to privacy concerns (withheld images). Participants were first briefed on the goal of \systemname{} and the study and were then shown a video walkthrough of the tool before beginning the tasks.

\textbf{Task:} Participants were then asked to use the tool on each of the four images they brought to the study. They were only required to load each image into \systemname{}, express their privacy concerns and/or sharing intent with that image, and have a look at the privacy risks surfaced; they were \textit{not} required to make any changes to their images. 

\textbf{Measurements:} Participants were asked to fill in a pre-task and post-task questionnaire for each image. These questionnaires focused on measuring the tool's effectiveness against our design requirements through Likert scale questions (see Appendix \ref{sec:survey-questions} for the full set of questions).
Generally, these scales comprised of attitudinal questions such as ``I feel that the tool understood my privacy concerns and sharing intent'' where participants had to rate agreement from a scale of ranging from 1 (strongly disagree) to 5 (strongly agree). Following best practices in questionnaire design, some of our questions were reverse-coded.After completing the tasks with all four images, participants were asked to fill in a final questionnaire, which aimed to measure how \systemname{} addressed the SPAF barriers. The final survey also had a section with the System Usability Scale (SUS) questionnaire to evaluate overall usability.

\textbf{Exit interview:} After participants used \systemname{} on all four of their images, we conducted a final semi-structured interview. The questions we asked participants were informed by their responses to the questionnaires they filled out for each image and centered around understanding whether the individual design requirements were met. For example, we asked questions like: ``In the survey you indicated that the tool helped/did not help you identify privacy risks you hadn't considered before. Could you share more about what led you to this conclusion?''. Participants were encouraged to refer to the four images they tested and give examples while answering these questions. We ended with a brief demographic questionnaire.  This mixed-methods approach allowed us to evaluate \systemname{} both quantitatively and qualitatively.

\textbf{Analysis:}  We employed a mixed-methods approach to analyze our data. The interview responses were thematically analyzed by two researchers individually who later came together to resolve any conflicts \cite{braun2006using}. Our approach combined both deductive and inductive coding. We began with a deductive coding frame to assess whether \systemname{} adhered to our five design requirements (DR1–5) and addressed the SPAF barriers (awareness, motivation, ability) \cite{das2022security}. In parallel, we remained open to emergent themes that reflected participants' unanticipated concerns, reactions, or values. These inductive insights revealed additional opportunities and limitations that were not captured by the original design requirements. 

To complement our qualitative analysis, we also analyzed the post- versus pre-task questionnaires participants filled out. First, outside of the SUS scale items, all reverse-coded items were recoded before analysis to ensure a consistent interpretation of scale direction (i.e., with a 1 indicating a negative impression, and a 5 indicating a positive impression). We then calculated descriptive statistics (means and standard deviations) for metrics related to each design requirement and SPAF barrier. To model the relationship between the use of \systemname{} to participants' perceived expression and privacy risk, we fit two random-intercept ordinal logistic regression models, accounting for repeated measures with a random-intercept term for participant ID.

\begin{figure*}[h]
    \centering
    \includegraphics[width=\textwidth]{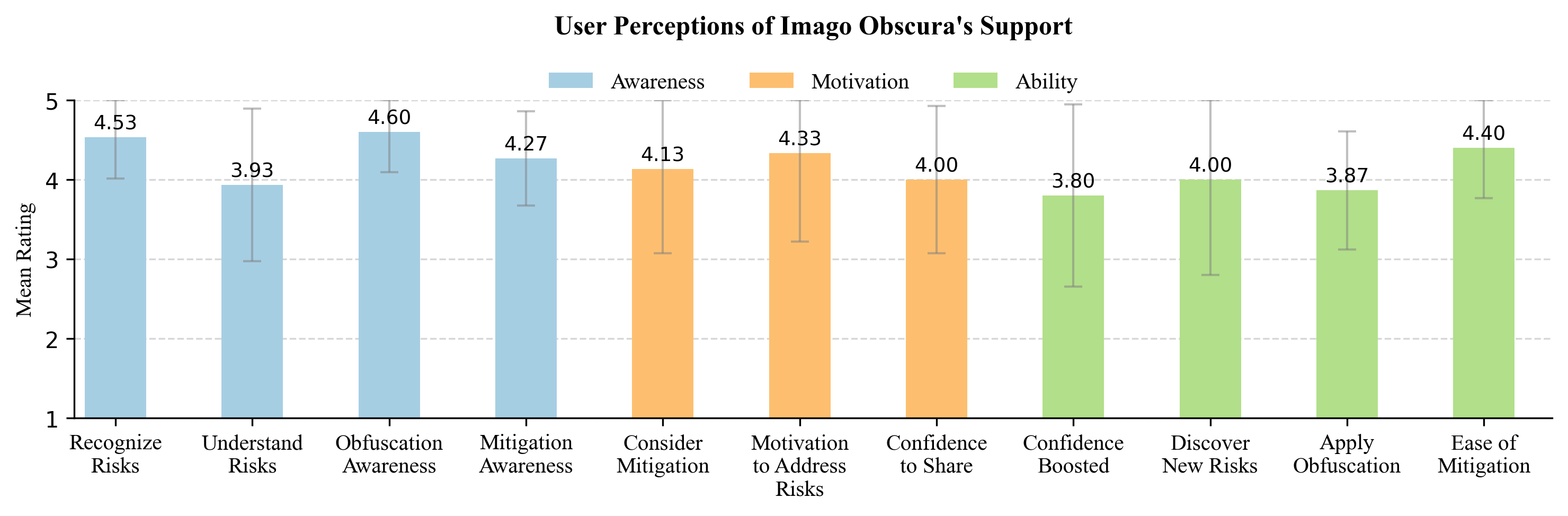}
    {\small\textit{Note:} Based on final questionnaire likert ratings (1 = strongly disagree to 5 = strongly agree); reverse-coded items were re-coded for consistency.}
    \vspace{-10pt}
    \caption{Participants rated \systemname{} highly across all three SPAF barriers—awareness, motivation, and ability—suggesting the tool effectively supports users in adopting pro-image privacy behaviors.}
    \Description{A bar chart showing mean user ratings for ten different statements grouped under three categories: awareness (blue), motivation (orange), and ability (green). Each bar includes a mean score and error bar. The highest-rated items are “Obfuscation Awareness” (4.60) and “Recognize Risks” (4.53), while “Confidence Boosted” (3.80) and “Apply Obfuscation” (3.87) received comparatively lower but still positive ratings. Ratings are based on a 5-point scale, with all means above 3.7, indicating overall positive responses. Error bars represent standard deviation.}
    \label{fig:spaf_chart}
\end{figure*}

\subsection{Findings}
Participants expressed a variety of sharing intents for the images they brought in --- from documenting personal experiences, celebrating achievements, highlighting casual moments, and showcasing scenic or humorous content with friends and followers. The privacy concerns they expressed about their \textit{withheld} images --- i.e., images they \textit{wanted} to share but did not for privacy reasons --- included: violating the privacy of others pictured without consent; unintentionally disclosing sensitive personal spaces or geographic locations; and, sharing content that could be misunderstood or pose reputational risks.

\subsubsection{\textbf{How does \systemname{} impact users' awareness of, motivation to address, and ability to address image privacy risks?}}

\paragraph{\textbf{Awareness Barrier}:}
 Participants strongly indicated that \systemname{} enhanced their awareness of privacy risks in images (M=4.53, SD=0.64) and improved their understanding of potential privacy risks (M=3.93, SD=0.96). P13 noted during the interview, ``It didn't occur to me that somebody might be able to identify the building, and that could be a privacy risk in a certain type of photo''. P12 mentioned ``I think definitely I would use something like [\systemname], in the future, if I'm taking a photo, in my house where you could see more of a layout,  especially being a small woman, I think more about personal safety and stuff''. Participants also reported becoming more aware of different obfuscation techniques available to address image privacy risks (M=4.60, SD=0.50). The tool successfully enhanced participants' awareness of how to address privacy risks in images (M=4.26, SD=0.59). P8 explained, ``it gave different suggestions for how you can replace it, like replacing the people with statues that look similar. It's not something that I would have thought of''.

\paragraph{\textbf{Motivation Barrier}:}
 Our results suggest that \systemname{} positively influenced participants' motivation to address image privacy risks. Participants reported that using the tool made them more likely to consider mitigating privacy risks in their images (M=4.13, SD=1.06).
 Participants also felt motivated to take steps to address privacy risks in the images they share online (M=4.33, SD=1.11). Furthermore, participants expressed increased confidence in their ability to share images while mitigating key privacy risks (M=4.00, SD=0.92). As P3 remarked: ``I started looking to make sure what could be a risk. And I think I didn't do that before, when I was even posting. I'm usually a very careful person, but it definitely helped me become, like, a little bit more aware of that.''
 
\paragraph{\textbf{Ability Barrier}:}
 The tool successfully supported participants in overcoming the ability barrier by making it easier for them to address privacy risks in their images (M=4.40, SD=0.63). Participants felt that the tool helped them identify risks in their images that they hadn't considered before (M=4.00, SD=1.19), although there was more variation in responses to this question compared to others. P8 explained, ``It did highlight almost all the privacy concerns I had, probably even uncovering some concerns which I did not think of, including geo tagging and so on.'' Participants also reported feeling confident about sharing their images online after using the tool (M=3.80, SD=1.14) and also felt supported in effectively applying techniques to mitigate pertinent privacy risks (M=3.8, SD=0.74). P12 stated: ``I think it made me feel more comfortable posting the photos that I didn't post, right? Being able to, like, pick and choose what I wanted out and cover what necessarily shouldn't be online.''

\subsubsection{\textbf{How well does \systemname{} adhere to design requirements?}}

\begin{figure*}[h]
    \centering
    \includegraphics[width=\textwidth]{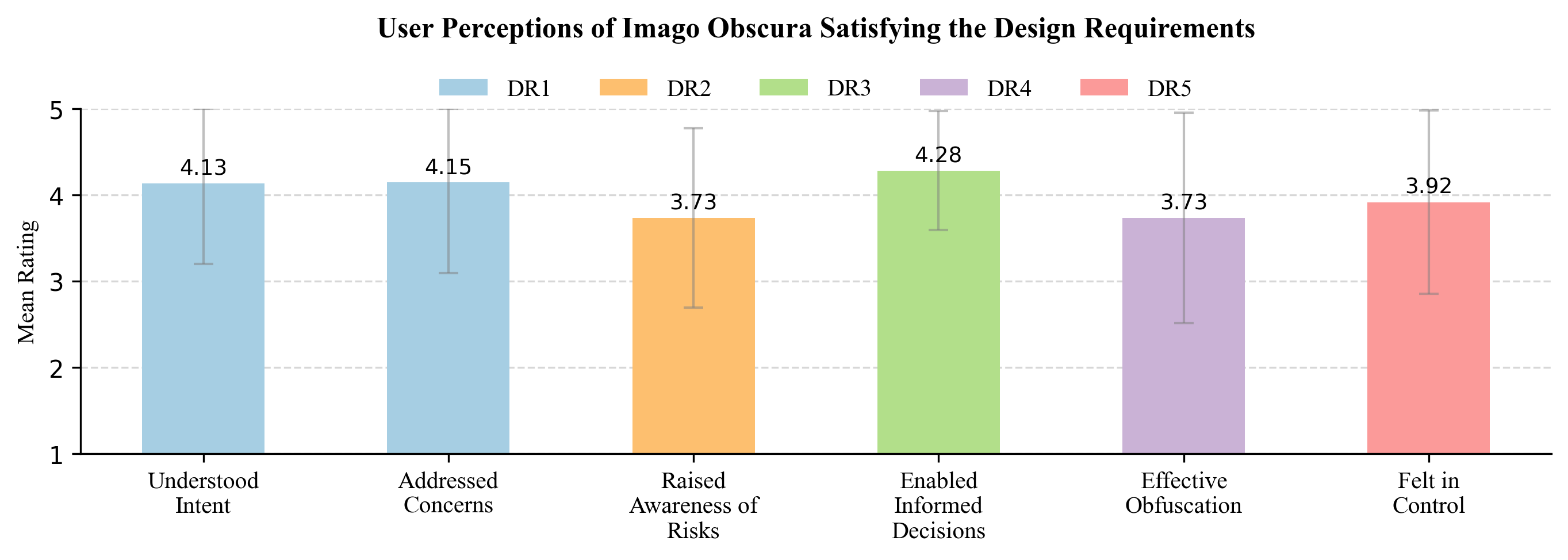}
    {\small\textit{Note:} Based on Post-tool use questionnaire likert ratings (1 = strongly disagree to 5 = strongly agree); reverse-coded items were re-coded for consistency.}
    \vspace{-10pt}
    \caption{Participants reported that \systemname{} satisfied all five design requirements (DR1–DR5).}
    \Description{A bar chart showing average user ratings for how well Imago Obscura satisfied five design requirements: DR1 (understood intent 4.13, addressed concerns 4.15), DR2 (raised awareness of risks 3.73), DR3 (enabled informed decisions 4.28), DR4 (effective obfuscation 3.73), and DR5 (felt in control 3.92). Ratings are based on a 5-point scale, with all means above 3.7, indicating overall positive responses. Error bars represent standard deviation.}
    \label{fig:dr-chart}
\end{figure*}

\paragraph{\textbf{DR1: Understands and Accounts for User-Articulated Privacy Concerns}}
Participants felt that \systemname{} understood their privacy concerns and sharing intent (M=4.13, SD=0.92) and effectively addressed their concerns (M=4.15, SD=1.05).
For example, P1 stated: ``[...] I had stated my privacy concerns, and the tool was able to pick up on that and also identify objects or people in the image that I had not considered obscuring before [...]'' More generally, users appreciated how the system recognized a broad spectrum of concerns, from identifying individuals in backgrounds to detecting revealing location information.

\begin{figure}[t]
\centering
    \centering
    \includegraphics[width=0.5\textwidth]{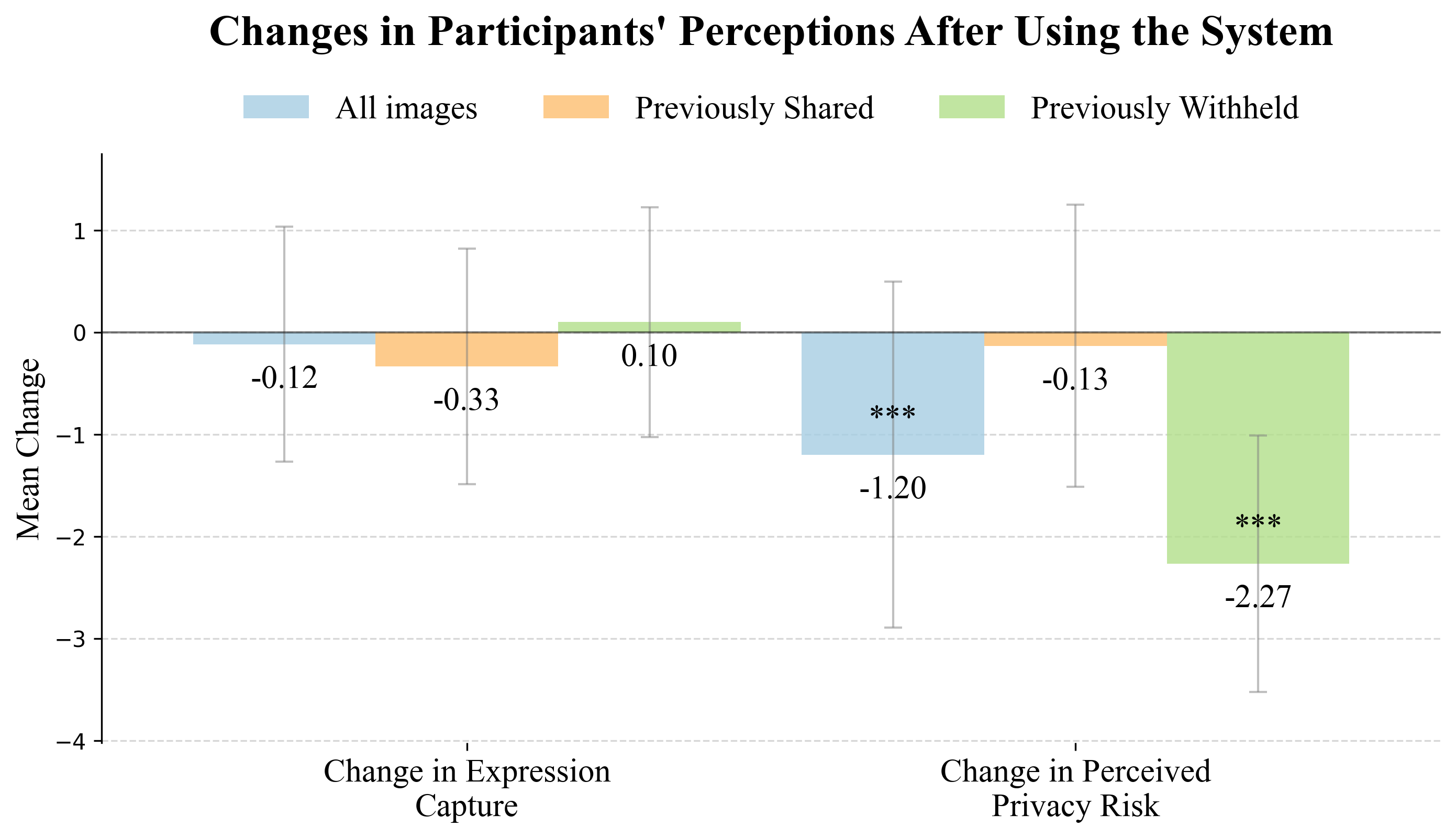}
    {\small{\textit{Significance}: $^{*}p<.05$; $^{**}p<.01$; $^{***}p<.001$} \\ \textit{Note}: Calculated by subtracting the post-tool use likert rating from the pre-tool use likert rating. A negative mean change indicates a decrease.}
    \caption{Using \systemname{} significantly reduces the perceived privacy risk in an image.}
    \Description{A bar chart showing average user ratings for how well Imago Obscura satisfied five design requirements: DR1 (understood intent 4.13, addressed concerns 4.15), DR2 (raised awareness of risks 3.73), DR3 (enabled informed decisions 4.28), DR4 (effective obfuscation 3.73), and DR5 (felt in control 3.92). Ratings are based on a 5-point scale, with all means above 3.7, indicating overall positive responses. Error bars represent standard deviation.}
    \label{fig:impact}
\end{figure}

\paragraph{\textbf{DR2: Expands Awareness of Content-level Privacy Risks}}
Many participants reported becoming more aware of privacy risks they hadn't previously considered (M=3.73, SD=1.04):
``I never really thought about specifics, like flags of places that I was visiting [...] People can look up the town and they could figure out everything.'' (P10)
The tool particularly heightened awareness around location leakage, the presence of bystanders, and background elements that could compromise privacy. ``...[I] actually have people in the background... I know one of them is specifically very obsessed with privacy issues. I have shared this photo before without blurring, but I think if I want to share it again now I will blur.'' (P15)
Accordingly, participants reported \systemname{} having an educational effect:  ``I think it's giving a pretty good taxonomy of risks with their rating on each picture... by running these four pictures, I kind of learned different ways of framings of some potential privacy risk and how they are ranked in the pictures'' (P14).
Similarly, P3 was surprised when the tool identified a logo that could reveal their location, noting it was ``something that I completely brushed over.'' Beyond physical locations, participants recognized subtle contextual identifiers in images: from visible phone screens (P7) to religious buildings (P8) and even flags (P9), which could reveal personal information. By highlighting overlooked risks, the tool broadened users' privacy awareness beyond their initial concerns.

\paragraph{\textbf{DR3: Empowers Informed Decision-Making}}

Participants reported feeling empowered to make informed decisions about addressing privacy risks in their images (M=4.28, SD=0.69). In particular, they appreciated that the tool explained both the identified risks and the corresponding obfuscation strategies, enabling them to weigh trade-offs. When asked what helped them decide which mitigation strategy to employ, a participant highlighted how ``when I hover the mouse... it showed you, like, a box'' and offered an ``explanation that what it does'', referring to how the tool visually highlighted areas of concern and clearly described the suggested mitigation strategy along with its relevant attributes.
Participants also appreciated the variety of obfuscation options provided, and that they could easily apply and remove the obfuscations to choose which they technique they preferred more.

Our quantitative results align with the finding that \systemname{} enabled more informed choices (Fig \ref{fig:impact}, Appendix Table \ref{tab:impact} ).  Across all images, participants' perceived privacy risks decreased significantly after using the tool (M=–1.2, SD=1.69; $\beta = -1.66, p < .001$ - \textit{statistically significant}). This effect was most pronounced for \textit{withheld} images where risk perceptions dropped by more than two points on average (M=-2.26, SD=1.26 $\beta = -4.43, p < 0.001$). In contrast, for shared images, the average reduction in perceived privacy risk was not significant (M=-0.13, SD=1.382 $\beta = -0.56, p = 0.285$) --- largely because participants did not harbor strong privacy concerns for these images in the first place. 

Moreover, we found no significant change in participants' belief that using \systemname{} to address privacy risks changed how well that image captured their sharing intent (M=–0.12, SD=1.15; $\beta = -0.24, p = 0 .518$). While absence of evidence cannot be considered evidence of absence, our findings at the very least suggest that \textit{if} the modifications introduced by \systemname{} are negatively impacting sharing intent, the effect is quite small. In sum, we can surmise that \systemname{} helped participants greatly reduce perceived privacy risk without compromising sharing intent --- especially for images that participants \textit{wanted} to share but withheld for privacy reasons. 

We identified four different scenarios that showcased how \systemname{} impacted users' decision-making on whether and how to mitigate image privacy risks:
\begin{enumerate}
    \item Users with no awareness of privacy risks in an image became aware of potential risks and took steps to mitigate them.
    \item Users aware but unconcerned about certain risks, upon receiving more information and mitigation options, made more confident decisions after weighing sharing intent against concerns.
    \item Users aware of specific privacy risks in an image found that the tool effectively identified and helped mitigate them.
    \item Users uncertain about the validity of their concerns gained clarity when the tool highlighted risk severity (i.e., low/medium/high), often increasing confidence in their decision to share or not share.
\end{enumerate}

\paragraph{\textbf{DR4: Facilitates application of obfuscation techniques}}
Participants rated the tool positively for its effectiveness at applying image obfuscation techniques (M=3.73, SD=1.21). The seamless integration of risk identification and mitigation capabilities received positive feedback, highlighting the tool's success in meeting DR4 by making privacy protection techniques accessible and effective. ``I think it was very intuitive, like you type and then you press generate, and then it gives you all the options, right? I think it's very clear'' (P5). Participants found that the tool made it easier to implement privacy protections that would otherwise require specialized skills:  ``it's like, more user friendly than Photoshop because of the AI part of it, and specifically that it specified things that you can change or you shouldn't change.'' (P4). Participants also mentioned that \systemname{} had a low learning curve, so that even users without technical backgrounds could apply sophisticated obfuscation techniques: ``The learning curve was very low, so it's pretty easy to learn.'' (P1).

\paragraph{\textbf{DR5: Ensures autonomy and granular control}}

\systemname{} successfully provided users with a sense of autonomy and control during the risk mitigation process (M=3.91, SD=1.06). Participants reported feeling in control while applying risk mitigation techniques, appreciating the ability to selectively address specific risks according to their preferences.

The system's approach of providing options rather than making unilateral changes was particularly valued: ``I felt in control, because I could, like, discard the changes or type like a more specific prompt if I wanted to, and still have, like, the autonomy to choose from, like the generated images [...] So I was able to go back in and select only like specific amount and generate based on that.'' (P3) This agency allowed users to carefully consider the privacy-publicity tradeoff for each image, weighing what elements were important to preserve sharing intent while mitigating potential privacy risks.

Many participants appreciated the object selection feature because it gave them granular control. `P10 stated: ``I was able to very clearly specify, like, I want to blur out this part of the image... being able to just click in... when I did click specific parts would usually get what I wanted, so being able to have more fine tuned control was a good feeling.''  This sentiment was echoed by others, with P3 explaining that precise selection helped them feel in control when iteratively refining obfuscations:  ``...it kind of altered my face. So I was able to go back in and select only like specific amount and generate based on that.'' Several participants (P6, P14, P15) requested even more advanced selection tools like those found in professional editing software, with P6 suggesting ``if there was something like a lasso tool or something that could be more flexible as compared to just, you know, explicit object selection in the image... that will be pretty useful.''

\subsubsection{\textbf{How usable is \systemname? \\}}

Overall, participants reported a positive usability experience, particularly in terms of ease of use and learnability. The usability evaluation revealed an average estimated SUS score of 70.1, indicating a good overall level of system usability. It is worth noting, however, that we accidentally omitted the system inconsistency question typically present in the SUS scale--- effectively reducing the maximum possible score to 90 from 100 (since we assume the most pessimistic case where all participants answered strongly agree to the question ``I thought there was too much inconsistency in this system.''). Thus the score we report above should not be compared against the standard SUS benchmarks. Instead, we focused more specifically on the individual items of the SUS. 

Participants demonstrated particularly high ratings for the system's ease of use (M = 4.2) and low perceived need for technical support (M = 4.47), suggesting a user-friendly interface. The system's learnability was also perceived positively, with users indicating they would quickly learn to use the system (M = 4.0). Conversely, the lowest-scoring areas included the perceived frequency of use (M = 3.47) and system integration (M = 3.87), which may warrant further investigation to enhance overall user engagement and system cohesiveness.

\subsubsection{\textbf{What other values, concerns, or reactions did participants express?}}

\paragraph{\textbf{Increases Social Context and Consent Considerations}}
Participants became more aware of location risks and bystander privacy after using \systemname.
P1, for example, noticed how the tool ``was able to pick up that there were cashiers and workers in the photos... that I had not considered obscuring.'' This recognition extended to interpersonal dynamics, as P7 discovered privacy concerns in captured interactions like a subject's hands touching another subjects shoulder that they hadn't previously considered problematic.
Another notable finding was that \systemname{} caused participants to reflect on consent in image sharing. P12 stated: 
``I don't like posting other people when I don't have their consent to do it,'' while P2 appreciated how the tool recognized that ``I need their permission to share it and their face.''
P14's experience exemplified this reflection on bystander consent, noting that even after previously sharing an image, they reconsidered an image as they remembered a subject from the photo was particular about privacy.

\paragraph{\textbf{Image Privacy Can Come at the ``Cost'' of Authenticity and Self-Presentation}}
Users face a fundamental tension between protecting privacy and maintaining authentic self-expression. Many participants discussed how image obfuscations could undermine the authenticity and communicative intent of their images. This tension was explicitly described by P14 who noted: ``I feel like it's really hard to balance... removing privacy concerns while maintaining the authenticity of the picture... that's really a trade-off from my perspective.'' Many participants rejected certain modifications that appeared artificial, with P12 explaining: ``I thought the avatar one... felt a lot more cheesy to me... if I'm editing a photo to get rid of something for a privacy concern, I don't necessarily want people to know that I had a privacy concern.'' This tension influenced sharing decisions, as P2 explained: ``I will not still use it... because I feel that the image itself will not have that spirit or will not share what that moment was.'' Some participants expressed a willingness to make minor privacy-enhancing edits but drew the line at modifications that fundamentally altered the image's meaning or appearance. P3 articulated this threshold clearly: ``If there are smaller things that really need to be blurred, I would use it for very small things, but for bigger things... I wouldn't use it... because it completely alters the entire setting that I'm in. And if it does that, then there's no point of posting the picture in the first place.''

\paragraph{\textbf{Improving the Quality of Generative Replacements Can Reduce the Authenticity ``Cost'' of Privacy}}
While privacy obfuscations sometimes came with a perceived authenticity ``cost'', participants found the quality of generative replacements helped them navigate this trade-off. Generative content replacement techniques were accepted only to the extent that the replaced content was realistic. P3 more broadly captured this sentiment: ``I think it makes it seem fake, which I don't want it to appear faked or masked, because then people know that, and like, they will probably be more curious to look into the image, right?''. However, participants also indicated that high-quality generative replacements could mitigate this trade-off: ``after replacing the roof, it does not make the picture look bad... if it's not making the picture look bad or look weird, I don't see any trade off.'' (P14). The quality of generative content replacement algorithms, thus, plays a critical role in determining whether users would adopt privacy-enhancing modifications. As these algorithms improve over time, we might expect more broad acceptance of privacy-preserving obfuscations.

\section{Discussion}
To summarize, we engaged in a user-centered design process to build an image privacy AI-copilot --- \systemname{} --- that enables users to identify and mitigate privacy risks in images that they seek to share online. Our findings suggest that \systemname{} appeared to strike an appropriate balance between raising awareness of privacy risks and affording users agency to address those risks in a manner that did not compromise their sharing intentions. Users felt more confident in their decision-making, whether or not they chose to mitigate risks. We nevertheless discovered a number of considerations, limitations, and opportunities for future work that we discuss in more detail here.

\subsection{Human-AI collaborative systems can overcome the limits of purely automated systems}

Reducing the role of the ``human-in-the-loop'' through automation has long been a top-level objective of the security community \cite{cranor2008framework}. The usable security community, in contrast, has pushed back against this narrative by outlining how automation has its limits \cite{edwards2008security} --- for example, all automated systems have failure conditions, and purely automated security systems can make handling failure cases even more difficult for users. Generative AI technologies --- for all the risks they bring to privacy and security \cite{lee2024deepfakes} --- provide an interesting new opportunity to create human-AI collaborative systems that overcome the limitations of purely automated approaches in helping users make security and privacy decisions without overwhelming them with choice.

For example, unlike traditional image-generation or obfuscation tools, our approach with \systemname{} focuses on enabling users to navigate the nuanced decisions involved in managing the privacy/publicity boundary of content in an image being shared online, rather than merely automating the obfuscation process. Indeed, our findings indicate that it is critical to raise users' awareness of image privacy risks, expose available mitigation strategies, and make clear the implications of each choice.
While prior research has emphasized automating privacy protection, our work demonstrates that keeping users in the loop allows for a deeper and more meaningful engagement with their privacy decisions. The copilot design strikes a balance between automation and manual control, enhancing user agency through informed decision-making and simplifying action.

\subsection{Generative AI privacy-copilots should be scaffolded with theory-informed prompting}

A key risk of incorporating generative AI into user-facing products is that large language models can hallucinate and be inaccurate \cite{kalai2024calibrated, marcus2020next}. While techniques --- like retrieval augmented generation \cite{lewis2020retrieval, karpukhin2020dense} --- have been proposed to reduce the likelihood of hallucinations in some task contexts, prior work has argued that ``hallucination'' may be an inevitable outcome of the stochasticity of large language models \cite{xu2024hallucination}. This context begs the question: if generative AI systems hallucinate, can they be ``trusted'' to help people make privacy and security decisions?

The approach we took in the development of \systemname{} was to use a scaffolded prompting process that grounded outputs in existing literature and theory on image privacy, usable security, and privacy. For example, we did not simply prompt GPT-4o to identify privacy risks in an input image --- we first identified and segmented the image, we provided user-specified concerns, and we provided a taxonomy of image privacy risks distilled from prior literature. This process constrained the output space of our model pipeline to reduce the likelihood of catastrophic or ungrounded hallucination, thereby increasing accuracy and robustness.

Indeed, our evaluation revealed that users valued the system's ability to highlight privacy risks specific to their concerns and sharing intent while also informing them of risks that they had not considered but could see the value in identifying. This approach, theory-informed prompting, can also help with quickly translating empirical findings into actionable design implications --- one can imagine, for example, improving the outputs of \systemname{} by having it distill new empirical findings from the usable privacy literature on image privacy risks as these findings emerge.

\subsection{Opportunities for improvement}
In the process of creating and evaluating \systemname{}, we identified opportunities that merit further consideration.

\subsubsection{Fostering agency, preventing over-reliance}
Our study revealed broader conceptual implications that warrant careful consideration. Participants exhibited contrasting reactions after using our tool, highlighting potentially unintended consequences. While some indicated they would likely share fewer images due to heightened risk awareness, others expressed increased confidence in sharing, believing all risks were adequately addressed. These opposing reactions point to a possible over-reliance on the tool: users who are more sensitive to privacy risks may use it to confirm their fears and self-censor; users who are less sensitive to privacy risks may feel like the tool definitively covers all their bases. While improving user confidence is generally positive, it's crucial to ensure users understand that residual risks may still exist. This observation underscores the delicate balance required between informing users of potential risks and inadvertently overemphasizing them. Future iterations should aim to strike this balance, empowering users with knowledge while encouraging critical thinking about privacy. 

\subsubsection{Guardrails to prevent malicious use}
While it is widely acknowledged that generative AI technology can be used maliciously \cite{gupta2023chatgpt} --- including in ways that exploit user privacy \cite{lee2024deepfakes, kelley2023there} --- our work seeks to shift the balance of power back to the end-users by leveraging the capabilities of these technologies to enable users to preserve their privacy more effectively.
We also recognize that the generative technology in \systemname{} could be misused, for example, to deceive viewers or spread misinformation. For instance, one participant wanted to replace their younger sibling with another person, while another chose to replace fast food on their plate with healthier alternatives. These examples may seem innocuous, but they raise concerns around consent and content integrity ---especially when manipulated content could mislead others or be taken out of context, leading to unintended consequences.

Our user study revealed that many participants were mindful of these ethical concerns and found creative ways to balance privacy protection with ethical responsibility. For example, the participant who initially wanted to replace their sibling with another person later opted to replace them with a pet. They felt this substitution was more plausible and less deceptive, making it a good compromise. This finding further motivates approaches proposed in prior work, which explored replacing sensitive content with similar, non-sensitive alternatives to preserve context \cite{xu2024examining, khamis2022deepfakes}.

In sum, there remains a need for future research to explore and evaluate both nudges and guardrails to ensure consensual uses of reference images, and to appropriately watermark generated content so that it is clear when an image has been altered. 

\subsection{Limitations}
It is important to note that our evaluation, while designed to be representative of real-world scenarios by using users' own images, was conducted in a simulated laboratory setting with a limited number of participants who are not representative of all people who share images online. As such, the results should be interpreted with appropriate caution. Despite this limitation, we believe the insights gained remain valid and offer interesting perspectives on user interaction with privacy-enhancing tools.

We note that for the purposes of this work, our threat model excluded the institutional privacy risks that emerge from using third-party AI model providers. While our participants did not directly express privacy concerns regarding the use of third-party models, we recognize that using these models will exclude users who do not trust third-party providers. There are a number of locally deployable vision-language models that can be integrated to help address these concerns, but we leave it to future work to explore trade-offs between use of local models and the quality / acceptability of the obfuscations they generate.

\section{Conclusion}
We present \systemname, an AI-powered image privacy copilot that helps users make more informed decisions when navigating the privacy/publicity boundary \cite{palen2003unpacking} in online image sharing. We distilled five concrete design requirements for our tool following a formative study with seven image manipulation experts. Based on these requirements, \systemname{} enables users to articulate their image-sharing intent and privacy concerns, surfaces contextually relevant privacy risks, and recommends appropriate obfuscation techniques to address privacy concerns while minimally compromising sharing intent. Our implementation integrates a pipeline of generative AI models into an image editing tool, scaffolded by a theory-grounded prompting approach that leverages prior literature on image privacy and usable security. Through a summative evaluation with 15 participants who tested \systemname{} on their own photos, we found that the system  enhanced participants' awareness of, motivation to address, and their ability to mitigate relevant privacy risks in images they wanted to share online. As a result, it improved participants' confidence that they understood and could address pertinent privacy risks in their images, and thus their ability to make informed decisions about whether or not to share an image with or without obfuscation. More generally, our findings demonstrate how the interactive capabilities of modern generative AI technologies can help strike an effective balance between the benefits of automation and manual control for technologies that aim to simplify end-user privacy decision-making.

\begin{acks}
This work was generously funded, in part, by NSF SaTC Grant \#231629. We also thank Isadora Krsek for her valuable suggestions in improving the figures, Hank Lee for his insights on the analysis in the paper, and Yuxuan Li, Pradyumna Shome, and Minjung Park for their thoughtful feedback.
\end{acks}

%TC:ignore 
\bibliographystyle{ACM-Reference-Format}
\bibliography{sample-base}

\twocolumn[\newpage]

\appendix
\section{Appendix}

\begin{table*}[h]
    \centering
    \resizebox{\textwidth}{!}{
    \begin{tabular}{|p{0.22\textwidth}|p{0.1\textwidth}|p{0.13\textwidth}|p{0.12\textwidth}|p{0.15\textwidth}|p{0.1\textwidth}|p{0.12\textwidth}|}
        \hline
        \textbf{Manipulation \newline Technique} & \textbf{Effectiveness \newline (Human Recognition)} & \textbf{Detectability} & \textbf{Visual \newline Harmony} & \textbf{Narrative \newline Coherence} & \textbf{Realism} & \textbf{Vulnerability} \\ \hline
        Masking/Colorfilling & High & Obvious & Weak & Low & Unnatural & Low \\ \hline
        Silhouette Masking & High & Obvious & Weak & Medium & Unnatural & Medium \\ \hline
        Blurring & Low & Obvious & Weak & High & Unnatural & High \\ \hline
        Pixelating & Low & Obvious & Weak & Medium & Unnatural & High \\ \hline
        Bar Replacement & High & Obvious & Weak & Medium & Unnatural & Low \\ \hline
        Point Light Replacement & High & Obvious & Weak & Medium & Unnatural & Low \\ \hline
        Cartoon Replacement & High & Obvious & Strong & High & Unnatural & Medium \\ \hline
        Inpainting/Removal & High & Subtle & Strong & Low & Realistic & Low \\ \hline
        Generative Content Replacement & High & Subtle & Strong & High & Realistic & Low \\ \hline
    \end{tabular}}
    \caption{Effectiveness Attributes of Image Obfuscation Techniques from Literature}
    \label{tab:technique_attributes}
\end{table*}

\subsection{Use case scenarios}\label{sec:usecase}
To further contextualize how we envision end-users might use \systemname{}, consider the following two common scenarios.

\subsubsection{\textbf{Scenario 1: Inadvertent Sharing of Sensitive Information}}
Users often share images online without careful consideration of pertinent risks, which can lead to the inadvertent disclosure of sensitive information \cite{henne2013awareness, kumar2013security, nyoni2018privacy, wang2011regretted}. 
In turn, these accidental disclosures can lead to, for example, embarrassment, regret, harassment, and job loss \cite{rashidi2018you,wang2011regretted,mantouvalou2019lost, vishwamitra2021towards}.

\textbf{Example:} Alice, a 25-year-old marketing executive, takes a selfie at her desk to share her excitement about a new project. She posts it on her public Instagram story without noticing that her computer screen in the background displays confidential client information.  A competitor sees the post, leading to a breach of client confidentiality. Alice receives a formal warning and nearly loses her job.

\subsubsection{\textbf{Scenario 2: Privacy Concerns Inhibiting Image Sharing}}
Users may want to share images online for informational and/or emotional support, but hesitate due to privacy concerns. As a result, the user might either self-censor entirely and miss out on accessing support they seek, or they attempt to find a workaround that is difficult to implement, ineffective at addressing the privacy concern, or diminishes their sharing intent \cite{li2022obfuscation}.

\textbf{Example:} Bob, a 40-year-old father, wants to share photos from a recent family vacation on Instagram so he can keep his extended family updated. However, he's concerned about his children's privacy and the potential for their images to be misused online. He considers several options: 1) not sharing the photos at all, missing out on connecting with friends and family, 2) spending hours manually editing each photo to blur his children's faces, which is time-consuming and diminishes the quality of the images. and 3) sharing only scenery photos without people, which fails to capture the family moments he wanted to share. Ultimately, Bob feels frustrated by the lack of an easy solution that balances his desire to share with his need for privacy.

\subsection{Curated list of obfuscation techniques} \label{subsec:obfuscation techniques}

Our tool enables the application of nine obfuscation techniques curated from existing literature \cite{li2017effectiveness, xu2024examining, khamis2022deepfakes}.

\label{sec:obf-lit}
\paragraph{Masking}
Replaces sensitive content with a solid box for complete obfuscation.

\paragraph{Silhouette Masking}
Replaces content with shapes, preserving context without revealing identity.

\paragraph{Blurring}
Softens details while retaining visual context, offering moderate privacy.

\paragraph{Pixelation}
Enlarges pixels to obscure details while keeping recognizable forms.

\paragraph{Bar Replacement}
Covers sensitive content with a thin bar, highlighting presence but hiding specifics.

\paragraph{Point-Light Replacement}
Uses dots to represent movement, preserving dynamics without revealing identities.

\paragraph{Removal}
Eliminates sensitive content entirely, filling in the background seamlessly.

\paragraph{Avatar Replacement}
Replaces individuals with avatars, maintaining social cues while protecting identity.

\paragraph{Generative Content Replacement}
Replaces sensitive elements with realistic alternatives, ensuring coherence.

\subsection{Formative Study Material}
\subsubsection{Task Material}
To help users in this process, we explained potential privacy threats and provide a list of sensitive content that may exist in images, which they could consider as they edit.
\begin{enumerate}
    \item Threats
    \begin{enumerate}
        \item Interpersonal Threats
        \begin{itemize}
            \item Threats within the social circle or specific people with whom they were connected on online social networks
            \item Threats outside the social circle or strangers with whom they were not directly connected
        \end{itemize}
        \item Institutional Threats
        \begin{itemize}
            \item Companies, including employees within companies
        \end{itemize}
    \end{enumerate}

    \item Sensitive Content
    \begin{enumerate}
        \item Personal Identification
        \begin{itemize}
            \item Faces and identities of individuals (including photo owner, family, friend, bystander)
            \item Personal documents (ID cards, passports, licenses)
            \item Contact information (addresses, phone numbers)
            \item Vehicle plates and identifying markers
        \end{itemize}

        \item Nudity and Sexuality
        \begin{itemize}
            \item Full or partial nudity
            \item Sexual content or suggestive poses
            \item Revealing or immodest clothing
        \end{itemize}

        \item Privacy and Personal Space
        \begin{itemize}
            \item Home interiors and private areas (bedrooms, bathrooms)
            \item Personal belongings and assets
            \item Screens displaying private information
            \item Unorganized or messy living spaces
        \end{itemize}

        \item Sensitive Personal Information
        \begin{itemize}
            \item Medical conditions and treatments
            \item Financial information (bank accounts, credit cards)
            \item Legal documents and sensitive printed materials
            \item Educational records
        \end{itemize}

        \item Behavioral and Social Content
        \begin{itemize}
            \item Alcohol consumption and party scenes
            \item Smoking and drug use
            \item Inappropriate or illegal activities
            \item Unprofessional behavior at work
        \end{itemize}

        \item Appearance and Self-Presentation
        \begin{itemize}
            \item Unflattering images or angles
            \item Embarrassing expressions or poses
            \item Grooming and sleep-related content
        \end{itemize}

        \item Location and Environmental Identifiers
        \begin{itemize}
            \item Specific locations or landmarks
            \item Event attendance (revealing time and place)
            \item Workplace or school environments
        \end{itemize}

        \item Relationships and Personal Moments
        \begin{itemize}
            \item Intimate or affectionate interactions
            \item Family gatherings or private events
            \item LGBTQ+ related content
        \end{itemize}

        \item Political, Religious, and Controversial Content
        \begin{itemize}
            \item Political affiliations or activities
            \item Religious symbols or practices
            \item Controversial texts or memes
        \end{itemize}

        \item Potential Safety Concerns
        \begin{itemize}
            \item Weapons (real or fake)
            \item Dangerous situations involving children or pets
            \item Accident scenes
        \end{itemize}

        \item Digital Privacy
        \begin{itemize}
            \item Screenshots of private conversations
            \item Social media content without permission
            \item Unauthorized photos of others
        \end{itemize}

        \item Miscellaneous Sensitive Content
        \begin{itemize}
            \item Food and dietary habits
            \item Pets and their behavior
            \item Personal interests and hobbies
            \item Low-quality or old photos
        \end{itemize}
    \end{enumerate}
\end{enumerate}

\subsubsection{Post-task Interview Questions}
Note: This component employed a semi-structured interview approach, with a pre-defined set of questions serving as a guide for the interviewer. The question bank covered various topics. However, not all questions were necessarily asked during each interview. The interviewer selected the most relevant questions based on the participant's responses and the available time, in order to maintain a focused and efficient interview process. The interview was rephrased and conducted in a conversational manner to ensure participants felt comfortable throughout the process.

\begin{enumerate}
    \item Participant Background and Experience
    \begin{enumerate}
        \item Could you briefly describe your background and experience with photo editing or graphic editing tools?
        \item Have you ever used image editing specifically for obfuscation? If so, can you describe your experience?
    \end{enumerate}

    \item Task Workflow and Thought Process
    \begin{enumerate}
        \item Could you walk us through your thought process while completing the task?
        \item Can you walk us through how you approached obfuscating one of the images?
    \end{enumerate}

    \item Techniques and Rationale Behind Choices
    \begin{enumerate}
        \item I noticed you used different techniques (e.g., blurring, pixelation, removal, generative content replacement). Can you explain why you chose this (refer to a technique) technique?
        \item Could you describe the techniques you used in the images? Why did you select these techniques?
        \item Can you explain why you chose one technique over another for a specific element of the image (e.g., why you used blurring instead of removal)?
    \end{enumerate}

    \item Obfuscation Tool Design and User Experience
    \begin{enumerate}
        \item What features or design elements would make the tool easier for you to use?
        \item What changes or improvements would make the tool easier to use for non-experts?
    \end{enumerate}

    \item Experience with Task Materials and Suggestions
    \begin{enumerate}
        \item What role, if any, did the examples and sensitive content lists play in your decision-making during the task?
        \item Did the list of sensitive content and threats influence your thinking during the task? If so, how?
    \end{enumerate}
\end{enumerate}

\subsection{Evaluation Study}
\subsubsection{Survey Questions} \label{sec:survey-questions}
\begin{enumerate}
    \item Pre Task
    \begin{enumerate}
        \item Image ID
        \item This image captures what I'm trying to share or express online.
        \item There are privacy risks in this image that would make me hesitate to share it online.
        \item I feel comfortable sharing this image online.
        \item Why and with whom would you like to share this image online?
        \item Could you describe what privacy concerns you have with this image, if any?
    \end{enumerate}

    \item Post Task
    \begin{enumerate}
        \item Image ID
        \item I feel comfortable sharing the original image online.
        \item There are privacy risks in this modified image that make me hesitate to share it online.
        \item This modified image captures what I'm trying to share or express online.
        \item I feel uncomfortable sharing this modified image online for reasons other than privacy.
        \item I feel that the tool understood my privacy concerns and sharing intent.
        \item I feel that the tool failed at addressing my concerns.
        \item I already knew about all of the privacy risks the tool showed me.
        \item I was able to make an informed decision about if and how to address privacy risks in my image.
        \item The tool failed to effectively apply image obfuscation techniques.
        \item I felt in control while applying risk mitigation techniques.
    \end{enumerate}

    \item SUS (1 Question was accidentally omitted)
    \begin{enumerate}
        \item I think that I would like to use this system frequently.
        \item I found the system unnecessarily complex.
        \item I thought the system was easy to use.
        \item I think that I would need the support of a technical person to be able to use this system.
        \item I found the various functions in this system were well integrated.
        \item I would imagine that most people would learn to use this system very quickly.
        \item I found the system very cumbersome to use.
        \item I felt very confident using the system.
        \item I needed to learn a lot of things before I could get going with this system.
    \end{enumerate}

    \item Final Questions
    \begin{enumerate}
        \item I feel that the tool helped me recognize privacy risks in the images I share online.
        \item The tool did not increase my understanding of potential privacy risks in my images.
        \item I feel more aware of the different obfuscation techniques available to address image privacy risks.
        \item The tool did not enhance my awareness of how to address privacy risks in images.
        \item Using this tool made me more likely to consider mitigating privacy risks in my images.
        \item I feel unmotivated to take steps to address privacy risks in images I share online.
        \item Using this tool, I feel confident that I can share images while mitigating key privacy risks.
        \item The tool made me feel more confident about sharing this image online.
        \item The tool did not help me identify risks in my images that I hadn't considered before.
        \item I feel that the tool supported me in effectively applying techniques to mitigate privacy risks.
        \item The tool did not make it easier for me to address privacy risks in my images.
    \end{enumerate}
\end{enumerate}

\subsubsection{Semi-structured Interview Questions}

\begin{enumerate}
    \item General Experience with the Tool
    \begin{enumerate}
        \item How would you describe your overall experience using Imago Obscura on the four images you brought to the study?
        \item Were there any features that stood out to you? Why?
        \item Were there any challenges you encountered while using the tool? If so, can you describe them?
        \item Can you recall the last time you attempted to obfuscate an image you shared online? How would you compare using Imago Obscura to this previous experience?
    \end{enumerate}

    \item Design Requirements
    \begin{enumerate}
        \item Looking at the survey responses for the four images, it seems you felt that the tool did/did not understand your privacy concerns and sharing intent. Could you elaborate on why you felt this way? Were there specific moments or features that influenced your experience? [DR1]
        \item You indicated that the tool helped/did not help you identify privacy risks you hadn't considered before. Could you share more about what led you to this conclusion? Were there specific risks that stood out or were overlooked? [DR2]
        \item In the survey, you mentioned that the tool did/did not help you make informed decisions about addressing privacy risks. Could you explain why you feel this way? Were there aspects of the tool that supported or hindered your decision-making process? [DR3]
        \item Your responses suggest that applying the image obfuscation techniques was easy/difficult. Can you describe your experience with this process? Were there specific parts that you found straightforward or challenging? [DR4]
        \item You noted that you did/did not feel in control while using the tool. Could you explain what contributed to this feeling? Were there features or interactions that enhanced or diminished your sense of control? [DR5]
    \end{enumerate}

    \item Other
    \begin{enumerate}
        \item I see you feel more/less comfortable sharing this image, after using the tool. Can you explain what led to this change in comfort? (ask for 2 images, if possible of opposing results)
        \item You indicated that the new image did/did not capture what you were trying to share or express online for image [Image ID]. Could you elaborate on how the modifications affected your ability to communicate your intent? 
        \item You mentioned that the tool made you feel more/less confident about sharing the image online. Can you explain why?
    \end{enumerate}
\end{enumerate}

\subsubsection{Demographic Survey}

\begin{enumerate}
    \item What is your age? [Number]
    \item What is your gender? [Options: Male, Female, Non-binary, Prefer not to say, Other (please specify)]
    \item How often do you post or send images to others? [Scale: 1 - Almost every day, 2 - A few times a week, 3 - Once a month, 4 - Rarely, 5 - Never]
    \item Have you used any image obfuscation techniques before? If yes, what forms or tools have you used? [Short answer]
\end{enumerate}

\subsection{Technical Evaluation} \label{sec:tech-eval}
 We focused our technical evaluation on the risk identification component because it informs the users subsequent actions. We qualitative assess the other components of our tool.

To assess the performance of \systemname's risk identification component, we conducted an evaluation using the DIPA2 dataset \cite{xu2024dipa2}. The DIPA2 dataset was released in 2024, and
provides object-level annotations of sensitive elements and their corresponding privacy risk category.
The granularity and recency of this dataset makes it
an ideal baseline for our evaluation.

\subsubsection{Dataset and Methodology}
We evaluated our model's performance on three attributes of the dataset which were relevant to \systemname{}:
\begin{enumerate}
    \item \textbf{Object sensitivity}: Identifying whether an object in the image may be a privacy risk (binary classification)
    \item \textbf{Risk category} Assessing the category of risk (multi-class classification, 0-5 categories, as defined by DIPA2 ---personal information, location of shooting, individual preferences/ pastimes, social circle, others' private/ confidential information or Other)
    \item {\textbf{Severity}} Determining the severity of the risk (High / Medium / Low). DIPA2 usees a 1–7 Likert scale for severity. However, for our tool, we adopted a more user-friendly representation by prompting the MLLM to predict High, Medium, or Low. Accordingly, we reduced DIPA2's baseline to a 1–3 scale to compare it with the output from our pipeline for analysis.
\end{enumerate}

\subsubsection{Results}
Table \ref{tab:performance} presents the performance of the model pipeline we use in \systemname{ } on these three tasks:

\begin{table}[H]
\centering
\begin{tabular}{lccc}
\hline
Task & Accuracy (\%) & Precision (\%) & Recall (\%) \\
\hline
Object sensitivity(binary) & 69.65 & 63.02 & 53.22 \\
Risk category(multi-class) & 82.93 & 16.48 & 57.05 \\
Severity(High/Med/Low) & 72.86 & - & - \\
\hline
\end{tabular}
\caption{\systemname's Risk Identification Component Performance}
\label{tab:performance}
\end{table}

\begin{table*}[h]
\centering
\begin{tabular}{p{4.5cm}lcccccc}
\hline
Measure & Image Type & Mean Change (SD) & $\beta$ & SE & $z$ & $p$ & Sig. \\

\hline
\multirow{3}{4.5cm}{Change in Expression Capture 

} 
& All images & -0.116 (1.151) & -0.235 & 0.364 & -0.646 & 0.518 & \\
& Previously Shared & -0.333 (1.154) & -0.427 & 0.556 & -0.769 & 0.442 & \\
& Previously Withheld & 0.1 (1.124) & 0.181 & 0.499 & 0.364 & 0.716 & \\
\hline
\multirow{3}{4.5cm}{
Change in Perceived Privacy Risk 

} 
& All images & -1.200 (1.695) & -1.665 & 0.364 & -4.566 & <.001 & *** \\
& Previously Shared & -0.133 (1.382) & -0.563 & 0.526 & -1.07 & 0.285 & \\
& Previously Withheld & -2.266 (1.257) & -4.428 & 0.823 & -5.379 & <.001 & *** \\
\hline
\multicolumn{8}{l}{Significance: $^{*}p<.05$; $^{**}p<.01$; $^{***}p<.001$} \\
\multicolumn{8}{p{17cm}}{Note: Significance and effect direction are derived from cumulative link mixed models (random-intercept ordinal logistic regression), accounting for repeated measures and participant-level variation.} \\
\end{tabular}
\caption{Changes in Participants' Perceptions Before and After Using the System}
\label{tab:impact}
\end{table*}

\subsection{MLLM Prompts}
\subsubsection{Image Privacy Risk Identification Prompt}-\label{sec:prompt-id} 
\vspace{0.5em}
\begin{lstlisting}[style=promptblock]
[Background]: You are an AI assistant with expertise in privacy and social media, tasked with protecting the user's privacy when sharing photos online, by identifying potential risks in a specific image and communicating them concisely and in non-technical language to the user.

[Goal]: Analyze the provided image and associated information to:

1. Understand the context of the image  
   * Examine the photo [image]  
   * Consider the user's purpose for sharing, if provided [text]  
   * Address user's privacy concerns, if any [text, image with green annotations]  
2. Identify potential sensitive content  
   * Refer to the Sensitive Content list [text list]  
   * Analyze all objects in the photo [text, annotated images, object list]  
3. Determine privacy risks based on steps 1 & 2  
   * Refer to common privacy risks in photo sharing [text list]  
   * Identify user's concern specific privacy risks, if any [text]  
4. For each risk, categorize its severity and specify potential threat actors

Your analysis will help you identify and communicate potential privacy risks to the user in a clear and actionable manner.

[MATERIALS] To achieve your goal, you have access to:

1. Primary Image [Original Image]  
   * The image the user wants to share  
2. User-Provided Context (optional)  
   * Sharing intent in the user's words [User Input]  
   * Privacy concerns expressed by the user  
     * Textual description in users words [User Input]  
     * Annotated image with concerns marked in green by the user [User Concern Region]  
3. Image Analysis [Pre-Scan Data]  
   * Visually annotated photo with red boxes marking all objects  
   * JSON dictionary of object annotations, including position, length, and width of bounding boxes  
4. Reference Materials  
   * Curated list of Potential Sensitive Elements  
   * Curated list of Potential Risks in sharing images online

Remember to prioritize user-provided privacy concerns when identifying risks and sensitive content.

[TASKS] Please follow these tasks to analyze the image and provide necessary privacy risk assessments:

1. Understand the Image Context:  
   1. analyze the image and the users sharing intent   
   2. Describe elements within green-bordered areas as user concerns (if present)  
   3. analyze all user concern (if provided)  
   4. Focus on specific elements, not general categories (e.g., "license plate" instead of "car")  
   5. use concise phrases for each element   
2. Identify Sensitive Elements  
   1. Reference the curated list of potential sensitive elements  
   2. Scan the entire image for sensitive elements  
   3. Scan the annotated image for sensitive elements  
   4. Scan the objects identified in the dictionary for potential sensitive elements  
   5. Include user-highlighted concerns as sensitive elements  
   6. Consider context-specific sensitive elements not in the curated list  
   7. When conducting analysis, first examine each object individually and assess it for sensitivity, and then analyze the relationships between objects in the image to identify potential sensitive information inferred in the image.  
   8. Combine similar elements to avoid duplicates. For example, "person 1", "person 2", and "person 3" can be combined as "person"  
3. Determine Privacy Risks  
   1. Identify potential privacy risks for each sensitive element  
   2. Refer to the curated list of potential privacy risks to identify risks present in the image that the user might have forgotten to consider  
   3. Combine the same risks which have different sensitive elements   
   4. Use clear, non-technical phrases (max 5 words per risk) Example: "Reveals personal information" instead of "Self Disclosure"  
4. Assess Each Privacy Risk   
   1. Categorize severity: High, Medium, or Low. If the risk contains elements marked by the user, prioritize those risks as high severity.  
   2. Specify potential threat actors (e.g., Public Users, Companies, Family/Friends)  
   3. List associated sensitive elements using concise phrases  
   4. Consider user intent and privacy concern: Ensure that the severity prediction accounts for the user's mentioned intent and specific privacy concerns.  
5. Ensure Comprehensive Coverage  
   1. All risks should be identified  
   2. Every sensitive element should have at least one associated privacy risk  
   3. All user concerns must be addressed in at least one privacy risk  
6. Review and Refine  
   1. Verify all tasks are completed thoroughly  
   2. Ensure clarity and consistency in assessments

CURATED LIST OF POTENTIAL SENSITIVE ELEMENTS

1. Identity and Personal Information  
   1. Person: Faces and identities of individuals (including photo owner, family members, children, friends, bystanders)  
   2. Identity: Personal documents (e.g., ID cards, passports, licenses), contact information (e.g., home address, phone numbers)  
   3. Place Identifier: Locations (e.g., home, workplace), scenery, or vacation spots that may be private  
   4. Vehicle Plate: Vehicle license plates and identifying markers  
2. Nudity and Sexual Content  
   1. Full or partial nudity or semi-nudity  
   2. Sexual content, suggestive poses, or erotic imagery  
   3. Revealing, immodest, or inappropriate clothing (e.g., swimsuits, underwear)  
3. Other People and Social Contexts  
   1. Person: Photos featuring others (e.g., family, friends, coworkers, bystanders)  
   2. Group events and social gatherings (e.g., parties, weddings)  
   3. Interactions with significant others or personal moments with others  
4. Embarrassing or Unorganized Environments  
   1. Table: Messy, unorganized, or cluttered home spaces (e.g., kitchen, living room, bathroom)  
   2. Unflattering grooming or sleeping shots  
   3. Low-quality or outdated photos that do not reflect the current state  
5. Violence and Criminal Activity  
   1. Weapon: Scenes depicting violence or harm (e.g., battlefield, firearms)  
   2. Criminal behavior or unlawful activities (e.g., drugs, vandalism, theft)  
   3. Dangerous objects (e.g., weapons, guns)  
6.  Medical and Health Conditions  
   1. Visible injuries, medical conditions, or medical treatments  
   2. Unflattering depictions of physical health (e.g., acne, wounds, bad teeth)  
   3. Photos taken during medical procedures or showing medical equipment  
7.  Alcohol, Drugs, and Partying  
   1. Cigarettes: Images showing drinking, smoking, or substance use  
   2. Social gatherings involving alcohol, drugs, or related paraphernalia  
   3. Partying or celebratory events with potentially controversial behaviors  
8.  Appearance, Grooming, and Physical Attributes  
   1. Cosmetics: Unflattering body features or grooming (e.g., messy hair, weight issues)  
   2. Clothing: Tattoos, piercings, or unusual fashion choices that may be controversial  
   3. Finger: Poses or expressions that reflect poorly on personal character  
9. Religious and Cultural Sensitivity  
   1. Religious symbols, clothing, or practices that might be sensitive  
   2. Cultural references or behaviors that could be misinterpreted or offensive  
   3. LGBTQ+ content that may be sensitive in certain contexts  
10.  Sensitive and Private Information  
    1. Screen: Screens displaying sensitive or personal information (e.g., emails, documents, monitor screens)  
    2. Printed Materials: Handwritten or printed details revealing personal or professional data  
    3. Unique or personal belongings that reveal too much about the owner  
11.  Illegal, Unlawful, or Copyrighted Content  
    1. Printed Materials: Images associated with illegal activities (e.g., drug use, piracy)  
    2. Content that might suggest unlawful behavior (e.g., trespassing, theft, vandalism)  
    3. Book: Copyrighted materials or unauthorized content (e.g., photos of artwork, copyrighted documents)  
12.   Politically and Socially Offensive Content  
    1. Printed Materials: Politically sensitive or controversial subjects (e.g., North Korean leader, racism memes)  
    2. Vulgar gestures, symbols, or language (e.g., middle finger, offensive memes)  
    3. Racism, hate speech, or other socially offensive materials  
13.   Personal Assets and Belongings  
    1. High-Value Assets: Cars, jewelry, antiques, art, and other valuable personal belongings  
    2. Pet: Photos of personal pets or animals that the individual owns  
    3. Electronic Devices: Personal electronics (e.g., laptops, phones)  
    4. Musical Instrument: Musical instruments and other personal items that might be sensitive to the owner  
14.   Factors Affecting Public Image and Reputation  
    1. Photo: Unflattering or embarrassing shots that may harm public perception (e.g., unflattering facial expressions, bad hair days)  
    2. Machine: Activities or settings that can be misinterpreted negatively (e.g., unorganized home, awkward social situations)  
    3. Old, poor-quality, or technically flawed photos that do not reflect current image  
15.   Food, Lifestyle, and Leisure  
    1. Food: Unhealthy or unappealing food (e.g., junk food, fast food)  
    2. Lifestyle: Overindulgence or gluttony in food or drink, smoking, cigars  
    3. Toy: Personal items such as toys that might reflect a certain lifestyle  
16.    No Need to Share or Irrelevant Content  
    1. Content irrelevant to the audience or context  
    2. Trivial or unnecessary details that don't add value to the viewer (e.g., insignificant events, mundane personal moments)

Although an object annotated image and an object dictionary is provided to help you identify sensitive elements, you should always add more sensitive elements if you find any. Identify as many sensitive elements as possible. If [User Concern Region] is provided, the elements in the green border should be considered as sensitive elements.

CURATED LIST OF POTENTIAL PRIVACY RISKS  
Based on the image, thoroughly go through each element in the image, does it look 

1. Self-Disclosure: Can we learn something personal or sensitive about the photo owner or subject from the content of the image?  
2. Identity Disclosure: Can we learn something personal or sensitive about the photo owner or subject from the content of the image?  
3. Sensitive Information Leakage: Does the image reveal any unintended or unauthorized confidential data about the photo owner or subject?  
4. Location Exposure: Can the image provide insight into the movements or locations of the photo owner or subject, potentially exposing their location?  
5. Bystander Disclosure: Does the image inadvertently reveal personal information about third parties, such as bystanders, potentially violating their privacy?  
6. Acquaintance Disclosure: Does the image expose personal information about individuals familiar with the photo owner or subject, raising privacy concerns?  
7. Any other privacy risks you can think of

   You can use these privacy risks as a reference to identify potential privacy risks. Combine the same risks which have different sensitive elements. Remember to use clear, easy to understand phrases (max 5 words per risk), that is instead of mentioning the risks as is, mention it in a way understandable to a non-technical user and specific to the context in less than a 6 word phrase. 

Here are easy to understand example phrases for each image privacy risk:

1. Self-Disclosure Risk

Examples:  
Risk: "Reveals personal details"  
Sensitive object: "Visible diary pages"

Risk: "Shows private habits"  
Sensitive object: "Medication bottles"

2. Identity Exposure Risk

Examples:  
Risk: "Reveals who you are" or "Reveals your identity"  
Sensitive object: "Face clearly visible"

Risk: "Shows identifying marks"  
Sensitive object: "Unique tattoo visible"

3. Confidential Information Leakage Risk

Examples:  
Risk: "Exposes private data"  
Sensitive object: "Computer screen contents"

Risk: "Reveals secret info"  
Sensitive object: "Visible document text"

4. Location Exposure Risk

Examples:  
Risk: "Reveals where you are"  
Sensitive object: "Landmark in background"

Risk: "Location can be inferred"  
Sensitive object: "Distinctive local architecture"

5. Bystander Risk

Examples:  
Risk: "Shows others nearby"  
Sensitive object: "People in background"

Risk: "Includes uninvolved persons"  
Sensitive object: "Stranger's face"

IMPORTANT NOTE:  
Always try to understand the context of the image, and keep that in mind.  
If the user has provided a sharing purpose, you should use it to get a deeper understanding of the image and identify the risks accordingly.  
Remember if the user has provided specific privacy concerns, you should address them first and then remember to add other privacy risks that you think are relevant to the image but not mentioned by the user.  
If the user has highlighted sensitive elements by green borders in the image, you should solve the privacy risks associated with those elements first. And then proceed with the other sensitive elements.

OUTPUT FORMAT  
Respond with a JSON array of privacy risk objects. Each privacy risk object should have the following structure:  
{  
  "privacy_risk_id": Unique ID for the privacy risk,  
  "privacyRisk": "In a easy to understand language phrase/explain the potential privacy-invasive risk in less than 5 words",  
  "severity": "High/Medium/Low",  
  "threatActors": ["ThreatActor1", "ThreatActor2", ...],  
  "sensitiveElements": [  
    {  
      "id": Unique ID for the sensitive element, // this should be unique for each sensitive element among all the sensitive elements in the image, and same sensitive element should have the same ID in all privacy risks  
      "element": "Sensitive element associated with this privacy risk", // use concise phrase to describe the sensitive element  
      "riskCause": "In a phrase explain why the sensitive element leads to the privacy risk?",
      "markedByUser": true/false // only if the [User Concern Region] is provided and the sensitive element is explicitly marked by the user through green borders, mark this as true, otherwise false
    },  
    ...  
    // ensure this list does not contain duplicates  
  ]  
}
\end{lstlisting}
\subsubsection{Image Obfuscation Recommendation Prompt}-\label{sec:prompt-rec}
\begin{lstlisting}[style=promptblock]
[Background]: You are an AI assistant with expertise in privacy and social media, tasked with protecting the user's privacy when sharing photos online, by recommending image manipulation/obfuscation techniques for specific sensitive elements and regions in image and communicating their attributes to the user concisely and in non-technical language to the user.

[Goal]: Your goal is to understand the context of the image and the user's sharing purpose first. And then recommend suitable obfuscation techniques for each identified sensitive element to protect the user's privacy.

Analyze the provided image and associated information to:

1. Understand the context of the image  
   1. Examine the photo [image]  
   2. Consider the user's purpose for sharing, if provided [text]  
   3. Consider the user's privacy concerns, if any [text, image with green annotations]  
2. Understand privacy risks & respective sensitive present in the image  
   1. Refer to the Privacy Risk identified in the image [text list]  
   2. Refer to the Sensitive Content Elements identified in the image [text list]  
3. Analyze the available image obfuscation techniques and their advantages and disadvantages  
   1. Refer to the available image obfuscation techniques [text list] and their attributes [text list]  
   2. Match it to the privacy risks based on your understanding of what is required by the image context and the identified privacy risks and sensitive elements  [text]

Your analysis will help you identify and recommend relevant image manipulation/obfuscation techniques and present attributes of the technique in a context specific manner understandable to non technical users.

[Materials]: To help you better understand the image and privacy risks, you will receive:

1. the original image [Original Image]  
2. user's privacy concern, if provided any text description [User Input] or annotated image highlighting the areas of concern in green [User Concern Region].  
3. a list of privacy risks and respective sensitive elements identified in the image [Identification Result]  
4. Reference Materials  
   1. Curated list of Available Image Obfuscation Technique  
   2. Curated list of Attributes of Each Image Obfuscation Technique

[Tasks]:   
Please follow these tasks to provide the necessary recommendations for the image:

For each sensitive element of each privacy risk identified, provide specific image manipulation technique recommendations to mitigate the privacy risk. To do so:

1. Understand the Image Context  
   1. Analyze the image, user's sharing intent, and user concerns  
   2. analyze the users sharing intent and user concern text and (green) annotated image (if provided)  
2. Determine Relevant Image Manipulation Techniques  
   1. For each sensitive element in an identified privacy risk refer to the curated list of image manipulation and the curated list of attributes  
3. Generate Recommendations  
   1. List suitable recommendations for each sensitive element (one manipulation type per recommendation)  
   2. Select up to 2 most appropriate recommendations per sensitive element  
   3. Provide 2-6 recommendations per privacy risk (mostly 2 x number of sensitive elements pointing to the privacy risk)  
   4. If the user has provided specific privacy concerns or preferences, you should ensure all user concerns have been addressed.  
   5. Be creative and prioritize aesthetics - so consider the generative replacement, dot representation, avatar replacement, and removal techniques prior to other techniques.  
4. Present Recommendation  
   1. Use context-specific, user-friendly phrasing  
   2. Analyze and present attributes to help users make informed decisions  
      1. Include equal amounts of advantages and disadvantages  
   3. Explain attributes in context-specific, understandable terms  
5. Ensure Comprehensive Coverage  
   1. Every sensitive element should have at least one recommended mitigation phrase suggesting an image manipulation technique  
   2. All user concerns must be addressed in at least 2 mitigation strategy recommendations  
6. Review and Refine  
   1. Verify all tasks are completed thoroughly  
   2. Ensure clarity and consistency in assessments

   

CURATED LIST OF AVAILABLE IMAGE MANIPULATION TECHNIQUE  
The obfuscation techniques can be in the types of:

* Generative Replacement: replace the sensitive element with a generative image.  
* Removal: remove the sensitive element from the image.  
* Dot Representation: use dots and lines to represent the sensitive element's pose or gesture. When showing the pose or gesture, prioritize the dot representation.  
* Avatar Replacement: replace the sensitive element with an avatar. When showing the pose or gesture, prioritize the dot representation. If you need to generate an avatar, please select this type instead of generative replacement. Avatar replacement is only suitable for faces, not the whole person.  
* Bar Replacement: replace the sensitive element with a bar.  
* Silhouette: replace the sensitive element with a silhouette.  
* Masking: mask the sensitive element with a rectangle.  
* Pixelating: pixelate the sensitive element.  
* Blurring: blur the sensitive element.

CURATED LIST OF ATTRIBUTES OF IMAGE OBFUSCATION TECHNIQUES

* GCR (Generative Replacement): High human recognition resistance, Subtle manipulation, Strong visual consistency, High contextual alignment, Realistic, Low reversibility risk  
* Inpainting/Removal: High human recognition resistance, Subtle manipulation, High visual consistency, Low contextual alignment, Realistic, Low reversibility risk  
* Masking/Colorfilling: High human recognition resistance, Obvious manipulation, Weak visual consistency, Low contextual alignment, Unnatural, Low reversibility risk  
* Bar Replacement: High human recognition resistance, Obvious manipulation, Weak visual consistency, Medium contextual alignment, Unnatural, Low reversibility risk  
* Point Light Replacement: High human recognition resistance, Obvious manipulation, Weak visual consistency, Medium contextual alignment, Unnatural, Low reversibility risk  
* Avatar Replacement: High human recognition resistance, Obvious manipulation, Weak visual consistency, High contextual alignment, Unnatural, Low reversibility risk  
* Silhouette Masking: High human recognition resistance, Obvious manipulation, Weak visual consistency, Medium contextual alignment, Unnatural, Medium reversibility risk  
* Blurring: Low human recognition resistance, Obvious manipulation, Weak visual consistency, High contextual alignment, Unnatural, High reversibility risk  
* Pixelating: Low human recognition resistance, Obvious manipulation, Weak visual consistency, Medium contextual alignment, Unnatural, High reversibility risk
Do not mention phrases about the complexity and time of the technique or the technical terms. 


Please consider these attributes that are more relevant to the image context and are more helpful for users to make informed decisions.

OUTPUT FORMAT  
Add recommendations to each privacy risk, and keep the privacy_risk_id the same. Ensure all provided privacy risks have at least one associated recommendation.  
Return the same JSON array structure as in the provided dictionary and follow the original order of privacy risks. Each privacy risk object should have the following structure:  
{  
  "privacy_risk_id": The same id as in the dictionary,  
  "recommendations": [  
    {  
      "element": the id of the sensitive element,  
      "manipulation_type": "Type of recommendation (Generative Replacement, Removal, Dot Representation, Avatar Replacement, Bar Replacement, Silhouette, Masking, Pixelating, Blurring)",  
      "type_description": "Use concise and natural non-technical language to describe the recommendation",  
      "prompt": "If the recommendation is Generative Replacement, provide a prompt for the stable diffusion model, describing what you want to generate. For other types, return an empty string.",  
      "advantages": ["Advantage1", "Advantage2", ...], // keep each advantage concise up to 5 words  
      "disadvantages": ["Disadvantage1", "Disadvantage2", ...] // keep each disadvantage concise up to 5 words  
    },  
    ...  
  ]  
}
\end{lstlisting}
%TC:endignore

\end{document}